\documentclass[fleqn,usenatbib]{mnras}

\usepackage{newtxtext,newtxmath}

\usepackage[T1]{fontenc}

\DeclareRobustCommand{\VAN}[3]{#2}
\let\VANthebibliography\thebibliography
\def\thebibliography{\DeclareRobustCommand{\VAN}[3]{##3}\VANthebibliography}

\usepackage{graphicx}	
\usepackage{amsmath}	

\title[Spectral and Timing Evolution of Mrk 1040]{Evolution of Accretion Properties in Mrk~1040 using long-term X-ray Observations}

\author[Nandi et al.] {
Prantik Nandi$^{1}$,\thanks{E-mail: prantiknandi007@gmail.com}
Narendranath Layek$^{2,3}$,
Sandip K Chakrabarti$^{1}$,
Sachindra Naik$^{2}$,
Priyadarshee P. Dash$^{2,3}$
\\
$^{1}$Indian Centre for Space Physics, 466, Barakhola, Netai Nagar, Kolkata 700099\\
$^{2}$Astronomy and Astrophysics Division, Physical Research Laboratory, Navrangpura, Ahmedabad - 380009, Gujarat, India\\
$^{3}$Indian Institute of Technology Gandhinagar, Palaj, Gandhinagar - 382055, Gujarat, India
}

\date{Accepted XXX. Received YYY; in original form ZZZ}

\pubyear{\the\year{}}

\begin{document}
\label{firstpage}
\pagerange{\pageref{firstpage}--\pageref{lastpage}}
\maketitle

\begin{abstract}
We present a comprehensive long-term, multi-epoch spectral and timing study of the Seyfert 1 Active Galactic Nucleus (AGN) Mrk~1040, utilizing X-ray observations spanning from 2009 to 2024 ($\sim$15 years). The source exhibits pronounced spectral and temporal variability, indicative of transitions between different accretion regimes in the vicinity of the central supermassive black hole. The earlier reported soft excess is re-examined within a uniform, physically motivated multi-epoch framework. We confirm the presence of this soft excess in the 2009 observation, where it is well described by a warm, extended Comptonizing corona with $kT_{\rm e,warm} \sim 0.26$~keV and a radial extent of $R_{\rm warm} \sim 30~r_g$. In subsequent epochs, the soft excess is not statistically significant, possibly due to a combination of enhanced ionized absorption, intrinsic weakening of the warm Comptonizing region, or partial truncation of the inner disc. A strong correlation between the soft and hard X-ray fluxes suggests a common physical origin for both components, likely within a multi-layered Comptonizing structure that evolved into a compact and thermally stable corona after 2013. The observed spectral variability, together with changes in the Fe~K$\alpha$ line strength, reflects the evolving coronal geometry and accretion flow dynamics. Variations in the intrinsic column density ($N_H$) further indicate that Mrk~1040 is embedded within a clumpy, dynamically variable absorber responding to changes in the accretion rate. Using the TCAF model, we estimate the black hole mass as $M_{\rm BH} = (4.50 \pm 1.62) \times 10^7~M_\odot$, consistent with previous estimates.

\end{abstract}




\section{Introduction}
\label{sec:intro}
Active Galactic Nuclei (AGNs) are primarily powered by the accretion of matter onto a supermassive black hole (SMBH) \citep{Rees1984}, typically with masses ranging from $10^5$ to $10^9$ M$_\odot$ \citep{KR1995, Peterson2004, EHT2019}. This accretion process generates significant radiation across the entire electromagnetic spectrum. In the case of AGNs, the standard accretion disk model \citep{SS73} predicts that the peak of thermal emission due to accretion appears in the optical/UV regime \citep{SM1989}. The X-ray spectrum of an AGN is primarily originated from inverse Compton scattering of these thermal photons by a hot ($T \sim 10^8$–$10^9$ K), optically thin electron cloud \citep{Haardt1991, Haardt1993}, often referred to as the Compton cloud, located near the central black hole \citep{ST80, ST85, CT95}. This mechanism produces a power law with a typical photon index of $\Gamma \sim 1.5$--$2.5$ \citep{Nandra1994, Reeves2000, Piconcelli2005, Page2005}, and an exponential cut-off at energies in the range of 100--300 keV \citep{ST80, Ricci2017, Tortosa2018}. These spectral features are directly influenced by the temperature, optical depth, and geometry of the Comptonizing region \citep{ST80, CT95}.

In addition to the primary power-law continuum, the AGN X-ray spectra often exhibit several secondary components. These include a reflection hump above 10 keV \citep{Ross1993, Ross2005, Garcia2013, Garcia2014}, low-energy absorption below 3 keV \citep{Risaliti1999}, and sometimes an excess emission in the soft X-ray band \citep{Halpern1984, Singh1985, Arnaud1985, Turner1989}. The hard X-ray photons produced by the Compton cloud illuminate the accretion disk, giving rise to a reflection spectrum, characterized by a broad hump in the 10-50 keV range \citep{Krolik1999} and fluorescent lines such as the Fe~K$\alpha$ line at $\sim$6.4 keV \citep{George1991, Matt1991}. In Seyfert~2 AGNs, a subclass of radio-quiet AGNs, the X-ray emission is often heavily absorbed by the neutral or ionized gas, resulting in a flatter observed spectrum \citep{Antonucci1993, Risaliti1999}. In contrast, many Seyfert~1 AGNs, which lack strong absorbers (column density $N_H \leq 10^{22}$ cm$^{-2}$, show a featureless excess emission below 3 keV, commonly known as the soft excess \citep{Halpern1984, Singh1985, Arnaud1985, Turner1989, Matzeu2020, Xu2021, Nandi2021, Jana2021, Nandi2023, Layek2025}. The origin of this soft excess remains debated, although leading interpretations include Comptonization of disk photons \citep{Czerny1987, Middleton2009, Done2012, Kubota2018, Petrucci2018, Petrucci2020, Nandi2021} or ionized reflection from the inner regions of the disk \citep{Fabian2002, Ross2005, Crummy2006, GC2010, Walton2013}. Considering the accretion dynamics and radiative processes around a black hole, the Two-Component Advective Flow (TCAF) \citep{Chakrabarti1989, Chakrabarti1990, CT95} model offers a self-consistent physical framework for explaining both the soft excess emission \citep{Nandi2021, Nandi2024} and the high-energy reflection hump \citep{CT95, Mondal2021} observed in X-ray spectra.

Markarian~1040 (Mrk~1040), also known as NGC~931, is a nearby Seyfert~1 AGN at a redshift of $z = 0.01667$ \citep{Huchra1999}. The mass of the central black hole is estimated as $\log(M_{\mathrm{BH}}/M_\odot) = 7.64 \pm 0.40$, determined by the stellar velocity dispersion method \citep{Zhou2010, De2013}, following the approach outlined by \citet{Nelson1995} and \citet{Tremaine2002}. The first X-ray observations of Mrk~1040 were conducted by ASCA in 1994, revealing a complex spectrum characterized by a warm absorber and a prominent broad fluorescent Fe~K$\alpha$ emission line. This line exhibited a full width at half-maximum (FWHM) ranging from approximately 16,000 to 70,000 km~s$^{-1}$ and an equivalent width of $550 \pm 250$ eV, suggesting emission from material in the vicinity of the central black hole \citep{Reynolds1995}. Subsequent XMM-Newton observations in 2009 showed the presence of a soft excess and soft X-ray time lags \citep{Tripathi2011}, and high-resolution Chandra HETG spectra at a different epoch revealed a multi-zone warm absorber \citep{Reeves2017}. While earlier work characterized the soft excess and identified the absorption features at individual epochs, a consistent long-term, multi-epoch study that applies physical accretion-flow models to track the evolution of the corona and the absorber across 2009–2024 has not been presented. In this work, we explore the long-term high-resolution and time-resolved X-ray spectral behavior of Mrk~1040 to better understand its AGN characteristics.

\section{OBSERVATION AND DATA REDUCTION}
\label{sec:obs}

\begin{table}
\centering
\caption{Log of observations of Mrk~1040.}
\label{tab:obs_log}
\setlength{\tabcolsep}{3.5pt}
\begin{tabular}{lllll}
\hline
\textbf{ID} & \textbf{Date} & \textbf{Obs. ID} & \textbf{Satellite/Ins.}  & \textbf{Exp.} \\
 & \textbf{(yyyy/mm/dd)} &  &   & \textbf{(ks)} \\
\hline
XMM1  & 2009-02-13 & 0554990101 & \textit{XMM}/EPIC-PN & 90.92 \\
& & & & \\
Su    & 2013-08-11 & 707046010  & \textit{Suzaku}/XIS+HXD & 137.44 \\
& & & & \\
XMM2  & 2015-08-13 & 0760530201 & \textit{XMM}/EPIC-PN & 94.00 \\
Nu1   & 2015-08-12 & 60101002002 & \textit{NuSTAR}/FPMA+B & 62.95 \\
& & & & \\
XMM3  & 2015-08-15 & 0760530201 & \textit{XMM}/EPIC-PN & 84.60 \\
Nu2   & 2015-08-15 & 60101002004 & \textit{NuSTAR}/FPMA+B & 64.24 \\
& & & & \\
XRTc  & 2015-08-13 & 00081532001 & \textit{Swift}/XRT & 5.20 \\
  & --2015-11-02 & --00037089003  &  & \\
& & & & \\
XRT21  & 2021-10-25 & 00014866001 & \textit{Swift}/XRT & 2.47 \\
& & & & \\
XRT24  & 2024-08-12 & 00014866002 & \textit{Swift}/XRT & 2.60 \\
\hline
\end{tabular}
\end{table}

Mrk~1040 has been observed by {\it XMM-Newton} \citep{Jansen2001}, {\it Suzaku} \citep{Mitsuda2007}, {\it Swift} \citep{Gehrels2004}, and {\it NuSTAR} \citep{Harrison2013} at different epochs. In this work, we used publicly available data from the HEASARC archive\footnote{\url{https://heasarc.gsfc.nasa.gov/cgi-bin/W3Browse/w3browse.pl}} and reduced and analyzed it using the \texttt{HEAsoft v6.30.1} package. The details of the observations are provided in Table~\ref{tab:obs_log}.

\subsection{\it XMM-Newton}
\label{sec:xmm}
Mrk~1040 was observed by {\it XMM-Newton} at three different epochs. The first observation (XMM1) was carried out in 2009 with an exposure time of 90.9 ks. This data was analyzed by \citet{Tripathi2011}, who reported a soft lag of approximately $10^4$ seconds. After a six-year observational gap, {\it XMM-Newton} carried out two additional observations (XMM2 and XMM3) in August 2015. The details of these observations are listed in Table~\ref{tab:obs_log}. In this work, we utilize data from the European Photon Imaging Camera (EPIC)/pn CCD \citep{Struder2001}. The Observation Data Files (ODFs) are processed using the Science Analysis System \texttt{SAS v18.0.0}\footnote{\url{https://www.cosmos.esa.int/web/xmm-newton/sas-threads}}, with calibration files updated as of 23 October 2019. We select only unflagged events (\texttt{FLAG == 0}) with \texttt{PATTERN $\leq$ 4}, corresponding to single and double pixel events. High background intervals due to soft proton flares are excluded by generating appropriate Good Time Interval (GTI) files to optimize the signal-to-noise ratio. For source extraction, we used an annular region centered on the source coordinates, with inner and outer radii of 5 and 30 arcseconds, respectively. To correct for possible photon pile-up, we generated {\tt epatplot} diagnostic curves for each observation and found that no pile-up was detected for the selection of annular region with the inner and outer radii of 5 and 30 arcsec, respectively. The background was extracted from a nearby circular region of 60 arcseconds radius, free from source contamination. Response matrix files (RMFs) and ancillary response files (ARFs) were generated using the \texttt{RMFGEN} and \texttt{ARFGEN} tasks, respectively. Finally, the spectra are grouped using the \texttt{GRPPHA} task with a minimum of 50 counts per bin in the 0.3--10.0 keV range.

\subsection{Suzaku}
\label{sec:suzaku}
{\it Suzaku} observed Mrk~1040 once on 11 August 2013 (Su) for an exposure of 137.44 ks (see Table~\ref{tab:obs_log}). In this work, we use data from both the X-ray Imaging Spectrometer (XIS) \citep{Koyama2007} and the Hard X-ray Detector (HXD) \citep{Takahashi2007}. Standard data reduction procedures are followed for both instruments, based on the official data reduction guide\footnote{\url{https://heasarc.gsfc.nasa.gov/docs/suzaku/analysis/abc/}} and using the latest calibration files (dated 2015-10-05)\footnote{\url{https://www.astro.isas.jaxa.jp/suzaku/caldb/}}. The event files are reprocessed using \texttt{FTOOLS v6.25}. For the XIS data, source spectra and light curves are extracted from a circular region of radius 200 arcseconds centered on the source coordinates. Background data are extracted from a nearby region on the same CCD chip using a circular region of 250 arcseconds. The final spectra and light curves are created by combining data from the two front-illuminated detectors (XIS0 and XIS3). Response files are generated using the \texttt{XISRESP} script. To avoid contamination from the Si K-edge in the {\it Suzaku} detectors, data in the 1.6--2.0 keV range are excluded. The spectra are grouped using the \texttt{GRPPHA} tool with a minimum of 200 counts per bin. For the {\it Suzaku}/HXD data, the unfiltered event files are processed using the standard pipeline. The output spectrum is generated using the \texttt{HXDPINXBPI} task. Corrections for non-X-ray background, cosmic X-ray background, and dead time are applied. The final HXD spectra are grouped using \texttt{GRPPHA} with a minimum of 1 count per bin.

\subsection{Swift}
\label{sec:xrt}
Mrk~1040 was observed by {\it Swift} on approximately 20 occasions between 2013 and 2024. In this work, we consider only data from the Swift X-ray Telescope (XRT; \citealt{Burrows2005}). During the period between September and October 2013, {\it Swift}/XRT observed the source six times. However, due to very low exposure (detectable only for 0.2 ks in total), these observations are not used in our analysis. As {\it Suzaku} also observed Mrk~1040 around a similar period, we utilize that data instead. After that, from August to November 2015, Swift/XRT observed the source seven times. Among them, two observations — on 13 August (XRT1) and 15 August 2015 (XRT2) — are nearly simultaneous with {\it XMM-Newton} and {\it NuSTAR} observations. Since no significant spectral variability is found throughout the 2015 observations, we combine all 2015 data (denoted XRTc) to increase the signal-to-noise ratio and study them separately. In 2016, Mrk~1040 was observed three times with XRT, but the source remained undetected. Similar non-detections are noted in 2018 and 2019, with only two short exposures of 0.2 ks and 0.9 ks, respectively. Finally, two more Swift/XRT observations were made on 25 October 2021 (XRT21, 2.47 ks) and 12 August 2024 (XRT24, 2.60 ks), during which the source was clearly detected. These observations are included in our study. The details of all usable Swift/XRT observations are provided in Table~\ref{tab:obs_log}. We extract spectra and light curves using the online tool `XRT product builder'\footnote{\url{https://swift.ac.uk/user_objects/}} \citep{Evans2009}, which performs all necessary calibrations and data processing. Spectral data are grouped using the \texttt{GRPPHA} task with a minimum of 10 counts per bin. The light curves are extracted in the 0.5--10.0 keV range.

\subsection{NuSTAR}
\label{sec:nustar}
{\it NuSTAR} consists of two identical focal plane modules, FPMA and FPMB, operating in the energy range of 3.0–79.0 keV. Mrk~1040 was observed nearly simultaneously with {\it XMM-Newton} and {\it Swift} on two occasions — 13 August 2015 (N1) and 15 August 2015 (N2), with exposures of approximately 62.95 ks and 64.24 ks, respectively. We include both N1 and N2 observations in our analysis. The details of the observations are provided in Table~\ref{tab:obs_log}. Standard data reduction procedures are followed to process the raw data from both modules. The \texttt{NUSTARDAS v2.1.2} software\footnote{\url{https://heasarc.gsfc.nasa.gov/docs/nustar/analysis/}}, along with the latest calibration files from CALDB\footnote{\url{https://heasarc.gsfc.nasa.gov/FTP/caldb/data/nustar/fpm/}}, is used to produce cleaned event files. The task \texttt{NUPRODUCTS} is used to extract source and background spectra and light curves. A circular region with 60 arcsecond radius centered on the source is used for extraction, and a circular region of 120 arcsecond radius is selected on the same detector for background, avoiding edges and source contamination. Background-subtracted light curves are generated with 500 s binning for both FPMA and FPMB. Response files are created using the \texttt{NUMKRMF} and \texttt{NUMKARF} tasks. Finally, the spectra are grouped using the \texttt{GRPPHA} task with a minimum of 50 counts per bin.

\section{TIMING ANALYSIS}
\label{sec:time}
The timing analysis of Mrk~1040 was conducted on the X-ray light curves obtained from observations with {\it XMM-Newton} and {\it NuSTAR} (see Table~\ref{tab:obs_log}). In our analysis, we used light curves binned at 500 s (0.5 ks). In order to study the cross-correlation and variability amplitude between different energy bands, we extracted the total light curves of {\it XMM-Newton} (0.3--10 keV) and {\it NuSTAR} (3--40 keV) observations across several energy bands. We initially focused on the 3--10 keV primary continuum, which likely originates from a geometrically thick, optically thin Comptonizing corona \citep{ST80, ST85}. To investigate the origin of the continuum photons in detail, we further divided this band into narrower sub-bands: 3--4, 4--5, and 5--6 keV. The Fe K$\alpha$ line dominates the 6--7 keV band, while 7--10 keV approaches the reflection-dominated regime. Below 3 keV, additional spectral features such as low-energy absorption and the soft X-ray excess are commonly observed. We studied this regime using the 0.3--0.5, 0.5--1.0, and 1.0--2.0 keV bands. To explore potential reflection features at high energies, we used broader energy bands above 10 keV:10--15, 15--20, 20--30, and 30--40 keV.

\subsection{Cross-correlation}
\label{sec:corr}
To understand the temporal behavior of Mrk~1040, we conducted a cross-correlation analysis between the light curves across various X-ray bands ranging from 0.3 to 40 keV. The correlation and time lag between these bands provide valuable insights into the spatial and temporal structure of the accretion flow and the corona. For this analysis, we employed two widely used cross-correlation techniques: the interpolated cross-correlation function (\texttt{ICF}; \citealt{Gaskell1987}) and the $\zeta$-transformed discrete cross-correlation function (\texttt{ZDCF}\footnote{\url{https://www.weizmann.ac.il/particle/tal/research-activities/software}}; \citealt{Alexander1997, Alexander2013}). The \texttt{ZDCF} method is particularly well-suited for the unevenly sampled and sparse data, and employs a binning strategy similar to that described by \citet{Edelson1988}. In our \texttt{ZDCF} analysis, we performed 12000 Monte Carlo simulations to estimate uncertainties in the correlation coefficients. We conducted a similar number of simulations to calculate the likelihood distributions for each \texttt{ZDCF}.

We start our analysis using the XMM1 observation. Previously \citet{Tripathi2011} reported a soft X-ray lag in this observation. The ICF and ZDCF analyses presented here place this result in a multi-epoch context, allowing a direct comparison of the coupling between different energy bands across subsequent observations. In XMM1 observation, we found a moderate correlation $(0.5 < \text{Peak}_\text{ZDCF} < 0.7)$ near zero lag ($\tau \sim 0.5\pm0.7$ ks) between the 3.0--4.0 keV and the soft X-ray bands below 3 keV. This suggests a tight coupling between the soft excess and the primary continuum during this epoch. However, this contribution appears to diminish in the 2--3 keV and 3--4 keV bands, with $\text{Peak}_\text{ZDCF}=0.70\pm0.05$ with a lag of $\tau=0.41\pm0.67$ ks. Within the primary continuum (3--10 keV), positive correlations are also observed across adjacent energy bands, although the correlation strength gradually decreases with increasing energy separation close to the zero lag. The decrease in correlation strength with increasing energy separation may suggest a variation of physical parameters, such as electron temperature and/or optical depth, inside the Compton cloud.

In the XMM2 observation, the soft X-ray band (0.3--3.0 keV) shows a significant correlation with the 3.0--4.0 keV range, with an average peak value of $\text{Peak}_\text{ZDCF} \geq 0.7$. This suggests that the soft excess component is either absent or weak during this epoch. A similar trend as seen in XMM1, i.e., a decrease in the correlation strength with increasing energy separation, is observed here, supporting a common variability origin. In the simultaneous Nu1 observation  with XMM2, no significant correlation is found between the primary continuum and energies above 10 keV, indicating a lack of short-timescale variability or reflection features in the high energy regime during this observation.

A similar trend is observed in the XMM3+Nu2 observation. A strong correlation is found at lower energies, with ($\text{Peak}_\text{ZDCF} \geq 0.7$), and a lag consistent with zero ($\tau \sim 0.1\pm1.0$ ks) is found in the lower energy bands, which gradually weakens with increasing energy. This indicates a weakening coherence between the energy bands. In XMM3, although the ZDCF formally reaches a higher apparent amplitude ($\sim0.65$) at around 25 ks, this feature lies at the edge of the temporal window, where the number of data pairs per bin declines sharply and the estimator becomes unstable. Such edge peaks are known to arise from windowing and sampling effects \citep{Edelson1988, Gaskell1987, Alexander1997, Alexander2013}. This bump is supported by a few bins with large uncertainties. It is not reproduced in adjacent energy-band pairs or in other epochs, and is far larger than the short near-zero lags typically observed in Seyfert inter-band correlations \citep{Uttley2005, Uttley2014}. We, therefore, adopt the lower but stable near-zero-lag peak when constructing the lag probability distribution, as it is based on many more data pairs, persists under rebinning, and is consistent across energy bands and epochs. However, no significant correlation is detected in the higher energy regime ($E > 10$ keV) from the Nu2 data.

\begin{figure*}
\begin{center}
\hspace*{-0.8 cm }
\includegraphics[trim={4 0.5cm 2cm 0},scale=0.82]{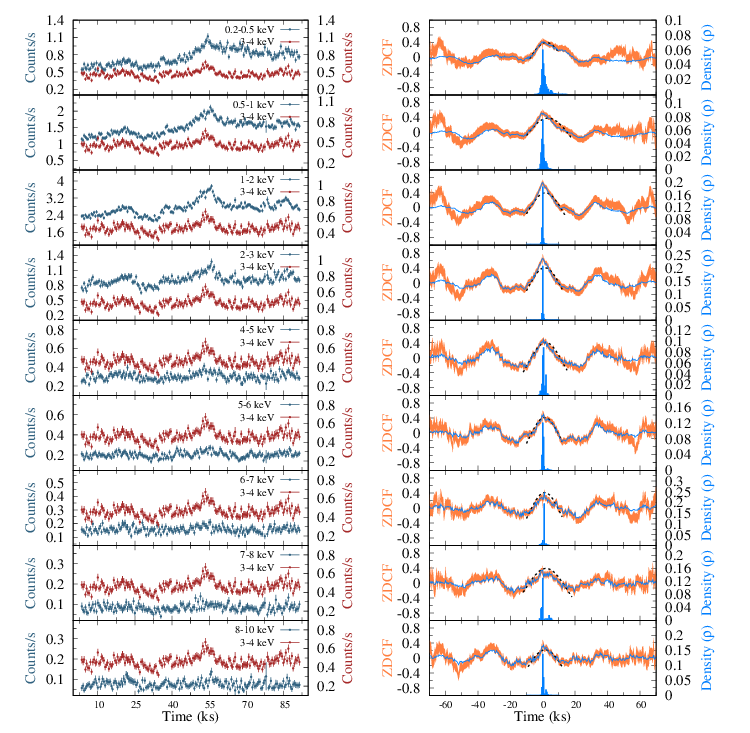}
\hspace*{0.8 cm }
\includegraphics[trim={4 0.5cm 2cm 0},scale=0.82]{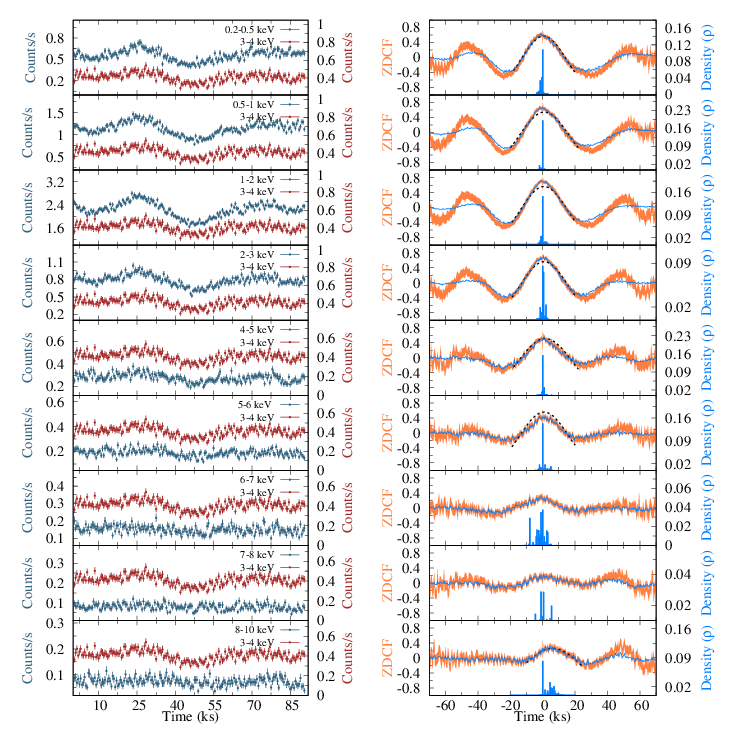}
\includegraphics[trim={4 0.5cm 2cm 0},scale=0.82]{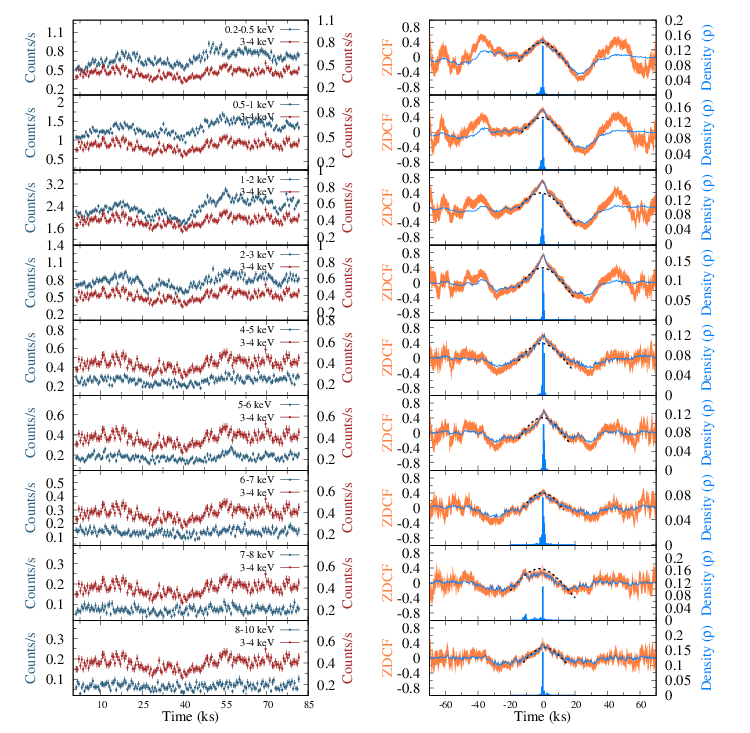}
\caption{The light curves and correlation functions (\texttt{ZDCF} and \texttt{ICF}) for different energy bands from the XMM1, XMM2, and XMM3 observations are shown. The likelihood distributions, simulated using 120000 points, are overplotted on the \texttt{ZDCF} curves.}
\label{fig:corr_xmm}
\end{center}
\end{figure*}

\begin{figure*}
\begin{center}
\hspace*{-3.5 cm}
\includegraphics[trim={4 0.5cm 2cm 0},scale=0.82]{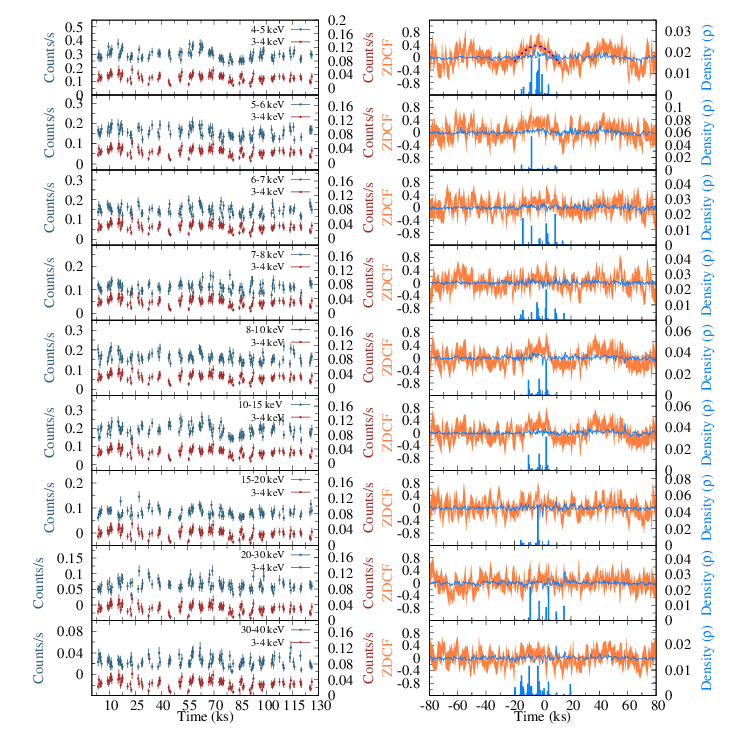}
\hspace*{0.6 cm}
\includegraphics[trim={2 0.5cm 5cm 0},scale=0.82]{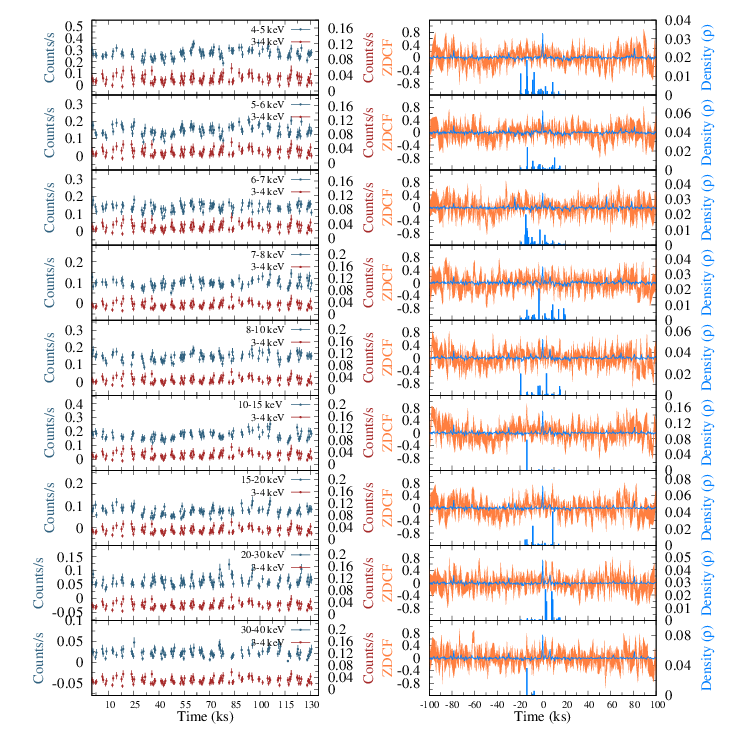}
\caption{The light curves and correlation functions (\texttt{ZDCF} and \texttt{ICF}) for different energy bands from the Nu1 and Nu2 observations are shown. The likelihood distributions, simulated using 120000 points, are overplotted on the \texttt{ZDCF} curves.}
\label{fig:corr_nu}
\end{center}
\end{figure*}

\subsection{Fractional Variability}
\label{sec:fvar}
\begin{figure}
\begin{center}
\includegraphics[trim={1 0 8cm 5},scale=1.75]{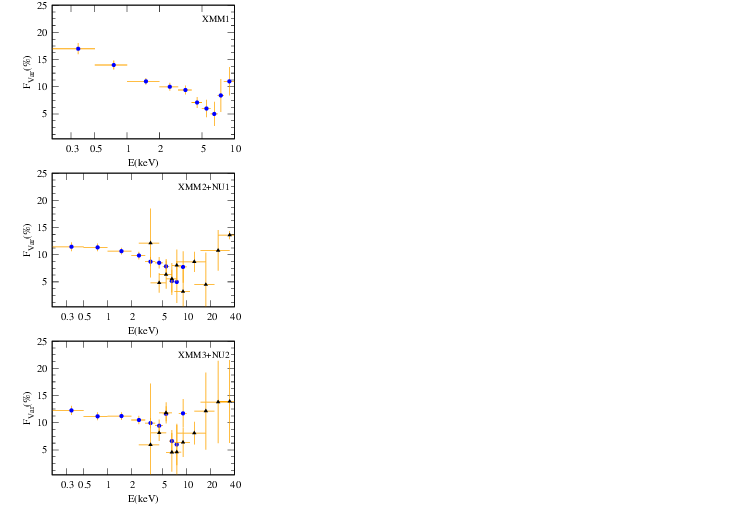} 
\caption{The energy-dependent fractional variability of Mrk~1040 in different epochs.}
\label{fig:fvar}
\end{center}
\end{figure}

X-ray emission from AGNs is known to exhibit variability over a wide range of timescales and energy bands \citep{Edelson1996, Nandra1997, Pascual1997, Edelson2012, Nandi2021, Nandi2023, Nandi2024, Layek2024, Layek2025b,Layek2025}. To quantify the variability in Mrk~1040 across different energy bands, we computed fractional variability amplitude, $F_{\rm var}$, which provides a normalized estimate of intrinsic variability relative to the mean count rate \citep{Vaughan2003}. It is defined as:

\begin{equation}
F_{\rm var} = \sqrt{\frac{\sigma^2_{\rm XS}}{\mu^2}},
\end{equation}
Where, $\mu$ is the mean count rate and $\sigma^2_{\rm XS}$ is the excess variance \citep{Nandra1997, Edelson2002}. The excess variance quantifies the intrinsic variance after subtracting the contribution from the measurement errors and is defined as:

\begin{equation}
\sigma^2_{\rm XS} = \sigma^2 - \frac{1}{N} \sum_{i=1}^{N} \sigma^2_i,
\end{equation}
where $\sigma^2$ is the total variance of the light curve, $\sigma_i$ is the uncertainty in the $i$-th count rate, and $N$ is the total number of data points. The normalized excess variance is defined as $\sigma^2_{\rm NXS} = \sigma^2_{\rm XS} / \mu^2$. The uncertainties in each variability parameter are estimated following the methods described in \citet{Vaughan2003} and \citet{Edelson2012}. The estimated values of the maximum, minimum, and mean count rates ($x_{\max}$, $x_{\min}$, and $\mu$), along with $\sigma^2_{\rm NXS}$ and $F_{\rm var}$, are provided in Table~\ref{tab:fvar}. The corresponding variation of $F_{\rm var}$ across energy bands is shown in Figure~\ref{fig:fvar}.

In the XMM1 observation, we observed a variability spectrum similar to that reported by \citet{Tripathi2011}. A higher level of variability is seen in the soft excess energy bands (0.3--3.0 keV) compared to the primary continuum bands (3.0--10.0 keV). Specifically, in the lower energy band (below 3 keV), the average fractional variability amplitude is $F_{\rm var} \sim (14.23 \pm 0.8)\%$, which is significantly higher than that of the primary continuum, where $F_{\rm var} \sim (7.08 \pm 1.9)\%$. Difference in the observed variability is interpreted as due to the possible presence of soft excess in Mrk~1040 during this observation, which is further discussed in Section~\ref{sec: pl}. The value of $F_{\rm var}$ for each band is presented in Table~\ref{tab:fvar} and the corresponding variation is also plotted in the upper panel of Figure~\ref{fig:fvar}. 

In the 2015 observations (XMM2+Nu1 and XMM3+Nu2), which are separated by only $\sim$2 days, we expected similar variability behavior. We found that Mrk~1040 did not exhibit significant variability during these epochs. The average fractional variability amplitude is $F_{\rm var} < 12\%$ across the 0.3–40.0 keV range. The variability is pronounced at lower energies, with $\sim 5\%$ near 8.0 keV, followed by a marginal increase above 8.0 keV in both XMM2 and XMM3 observations (see Table~\ref{tab:fvar} and Figure~\ref{fig:fvar}). For the {\it NuSTAR} data, the associated uncertainties are higher compared to the {\it XMM-Newton} observations (Figure~\ref{fig:corr_nu}), resulting in larger errors on $F_{\rm var}$ (Table~\ref{tab:fvar} and Figure~\ref{fig:fvar}). As a result, it is difficult to identify any clear trend in the energy dependence of $F_{\rm var}$ in the Nu1 and Nu2 data. The average $F_{\rm var}$ is $6.81\pm3.79\%$ for Nu1 and $8.91\pm 4.03\%$ for Nu2, both remaining below the 10\% level. 

The overall low variability, particularly in the hard X-ray band, suggests a relatively stable coronal structure \citep{Chitnis2009} during these observations. The increased variability in the low-energy bands may originate from fluctuations in the soft excess, likely associated with the inner accretion disc or a warm Comptonization region \citep{Petrucci2018}. The lack of strong variability further supports a scenario in which the accretion flow and coronal emission remain steady over the short timescale between the two observations.

The timing analysis reveals pronounced variability during the 2009 observation, coinciding with the observation of a strong, soft excess component \citep{Tripathi2011}.  In contrast, comparable variability is not observed in subsequent epochs, suggesting that the soft excess probably originates from a dynamic and compact region, plausibly associated with a warm Comptonizing corona \citep{Crummy2006, Done2012}.  Temporal variations in the soft excess are often linked to rapid fluctuations in the accretion rate or in the reprocessing geometry near the inner regions of the accretion flow \citep{Mehdipour2011, Petrucci2013}. These results support a scenario in which the presence of the soft excess contributes significantly to the observed variability, indicating the connection between the spectral state and the temporal characteristics of the source. To better understand the origin and evolution of these spectral components, we present a detailed spectral analysis of Mrk~1040 in the following sections.

\section{SPECTRAL ANALYSIS}
\label{sec:spec}
To explore the spectral properties of Mrk~1040 over a period of approximately 15 years (2009–2024), we utilize data from {\it XMM-Newton}/EPIC-pn, {\it Suzaku}, {\it Swift}/XRT, and {\it NuSTAR}. The spectral analysis is carried out using \texttt{XSPEC v12.14.1} \citep{Arnaud1996}. The likelihoods of the model are evaluated using $\chi^2$ statistics, and all the best-fit parameters are reported in the rest frame of Mrk~1040. Uncertainties on model parameters are estimated using the \texttt{error}\footnote{\url{https://heasarc.gsfc.nasa.gov/xanadu/xspec/manual/node79.html}} and Monte Carlo Markov Chain (MCMC)\footnote{\url{https://heasarc.gsfc.nasa.gov/xanadu/xspec/manual/node43.html}} technique implemented within \texttt{XSPEC}, with quoted errors corresponding to 90\% confidence intervals.

For the {\it NuSTAR} observations, we use the data up to 40~keV; beyond this energy range, the background dominates over the source signal and is therefore excluded from the analysis.In addition, the known bad channels are ignored during the analysis. For observations with low count rates, such as from {\it Swift}/XRT and {\it Suzaku}/HXD observations, we employ Cash statistics \citep{Cash1979} instead of $\chi^2$, as it is appropriate for low-count data. For all other observations with sufficient counts ($\geq$10 counts per bin), $\chi^2$ statistics are used to evaluate the goodness of fit.  

In this work, we adopt the following cosmological parameters: $H_0 = 70$ km s$^{-1}$ Mpc$^{-1}$, $\Lambda_0 = 0.73$, and $\Omega_M = 0.27$ \citep{Bennett2003}. Based on these values, the luminosity distance to Mrk~1040 is calculated to be 71.5 Mpc. The unabsorbed X-ray luminosity in different energy bands is estimated using the \texttt{clumin}\footnote{\url{https://heasarc.gsfc.nasa.gov/docs/xanadu/xspec/manual/node294.html}} task in \texttt{XSPEC}, applied to the \texttt{powerlaw}\footnote{\url{https://heasarc.gsfc.nasa.gov/docs/xanadu/xspec/manual/node163.html}} model with a redshift of $z = 0.01667$ \citep{Huchra1999}.

\subsection{Characterization of X-ray Spectrum}
\label{sec:characterization}
The X-ray spectra of Mrk~1040 were analyzed up to 10 keV by \citet{Tripathi2011}. However, the spectral properties of the source have not been explored in a broad energy range to date. To address this, we initially applied phenomenological models to characterize the source spectra in the 0.3--40 keV range across different observations. These models are later replaced with more sophisticated physical models to gain a deeper insight into the physical properties of the source.

\begin{figure*}
\begin{center}
\includegraphics[width=0.90\textwidth]{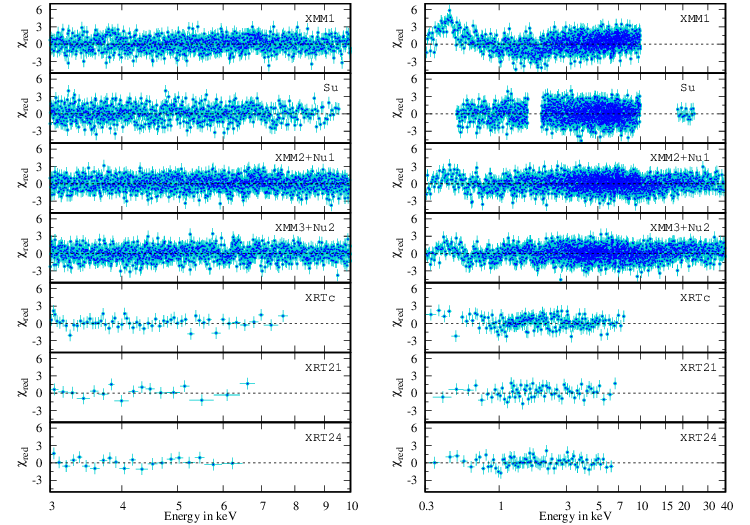}
\caption{{\it Left:} The variation of $\chi^2$ for the primary continuum using the model: \texttt{TBabs$\times$Constant$\times$(Powerlaw+zGauss)}. {\it Right:} The variation of $\chi^2$ over the full energy range using same baseline model. The presence of soft excess is clearly observed in the XMM1 observation (2009). However, this component is not found in any of the other observations across our $\sim$15-year observational period, from 2009 to 2024.}
\label{fig:chi}
\end{center}
\end{figure*}

We begin the spectral fitting by modeling the primary continuum, focusing on the 3.0--10.0 keV energy range. Current understanding suggests that the X-ray continuum originates from inverse Compton scattering of thermal photons from the accretion disk \citep{SS73} by a hot Compton cloud \citep{ST80, ST85}. As this is a non-thermal process, it results in a power-law-shaped spectrum \citep{ST80}. Consequently, we use the \texttt{Powerlaw} model to fit the primary continuum for all observations. In addition, we account for Galactic absorption using the \texttt{TBabs} model \citep{Wilms2000}, incorporating the Galactic line-of-sight hydrogen column density $(N_\text{H,Gal})$\footnote{\url{https://heasarc.gsfc.nasa.gov/cgi-bin/Tools/w3nh/w3nh.pl}}. For this work, we adopt a value of $N_\text{H,Gal} = 6.73 \times 10^{20}$ cm$^{-2}$ \citep{Kalberla2005, HI4PI2016}. During spectral fitting of the primary continuum, we find prominent positive residuals in the 6--7 keV range for the {\it XMM-Newton}, {\it Suzaku}, and {\it NuSTAR} observations, indicating the presence of an Fe~K$\alpha$ emission line. We modeled this line using a Gaussian component (\texttt{zGauss})\footnote{\url{https://heasarc.gsfc.nasa.gov/xanadu/xspec/manual/node131.html}}. Due to the low exposure time and limited spectral resolution, the Fe~K$\alpha$ feature is not detectable in the {\it Swift}/XRT observations. Therefore, for modeling the primary continuum, the baseline model becomes: \texttt{TBabs$\times$Constant$\times$(Powerlaw+zGauss)}. Here, the \texttt{Constant} component accounts for the cross-normalization between different instruments during the broadband spectral fitting. The ratio plots between models and the data are presented in the left panel of Figure~\ref{fig:chi}.

After fitting the primary continuum, we extended the source spectra to the lower energy domain (below 3 keV). The low-energy soft X-ray photons are primarily absorbed by the intrinsic hydrogen column density ($N_{H}$) along the line of sight. To account for this absorption, we included the multiplicative model \texttt{zTBabs} in our baseline model.

Additionally, we included a second power-law component to characterize the soft excess (if present) below 3 keV. Therefore, the baseline model for the spectral fitting in the 0.3–10.0 keV energy range becomes: \texttt{TBabs $\times$ zTBabs $\times$ Constant $\times$ (Powerlaw + zGauss + Powerlaw)}. As in \citet{Tripathi2011}, the soft excess component is also detected during our spectral fitting of XMM1 data. To determine whether the soft X-ray residuals arise from ionized absorption, we use a single power law model with an ionized absorber, ${\tt TBabs\times zxipcf\times(Powerlaw+zGauss)}$, yielding $\chi^2=2789.8$ for 1607 degrees of freedom (dof). Adding a soft power-law improved the fit to $\chi^2=2012.5$ for 1605 dof, with $F=256.15$ from an F-test, indicating a highly significant improvement. The absorber alone, therefore, cannot reproduce the soft curvature in XMM1, confirming that a distinct soft-excess component is required. However, no such excess was observed in the remaining datasets. In those cases, the primary continuum alone sufficiently described the low-energy spectra. To verify this, we performed an F-test, which indicated that adding an extra component did not significantly improve the spectral fits, confirming the absence of a significant soft excess in those observations.

Furthermore, absorption edges were identified near $\sim$0.6 keV and $\sim$0.8 keV in the \textit{XMM-Newton} spectra, while only one edge near $\sim$0.6 keV was present in the \textit{Suzaku} data. These features are consistent with those reported by \citet{Tripathi2011} for the XMM1 observation. To account for these spectral features, we introduced two \texttt{Edge}\footnote{\url{https://heasarc.gsfc.nasa.gov/xanadu/xspec/manual/node252.html}} components into the baseline model. The final model thus becomes: 
\newline
\texttt{TBabs$\times$zTBabs$\times$Constant$\times$(Powerlaw+zGauss+Powerlaw)\\$\times$Edge1$\times$Edge2}.

\begin{table*}
\tiny
\centering
\setlength{\tabcolsep}{2.0pt}
 \renewcommand{\arraystretch}{1.3}
\caption{The best-fit parameters of the baseline phenomenological model \texttt{TBabs$\times$zTBabs$\times$Constant$\times$(Powerlaw+zGauss+Powerlaw)$\times$Edge1$\times$Edge2} for the Mrk~1040 observations. The soft excess (SE) and power-law continuum (PC) luminosity are calculated in the energy range 0.3--40 keV. }
\label{tab:powerlaw}
\begin{tabular}{lccccccccccccccc}
\hline
ID & $N_\text{H}$   & $\Gamma_{\mathrm{PC}}$ & Norm$^\dagger_{\mathrm{PC}}$ & $\log L_{\mathrm{PC}}$ & $\Gamma_{\mathrm{SE}}$ & Norm$^\dagger_{\mathrm{SE}}$ & $\log L_{\mathrm{SE}}$ & Fe K$\alpha$ & EW &$E_{1}$&$\tau_{1}$&$E_{2}$&$\tau_{2}$& $\chi^2/\mathrm{dof}$ \\
  & ($10^{21}$ cm$^{-2}$) &                         & ($10^{-3}$)                  & (log [erg s$^{-1}$])   &                          & ($10^{-3}$)                  & (log [erg s$^{-1}$])   & (keV) & (eV) &(keV) & &(keV)&                       \\
\hline
XMM1 &  $1.70^{+0.76}_{-0.72}$ & $1.73^{+0.02}_{-0.02}$ & $7.18^{+0.16}_{-0.17}$ & $43.42^{+0.05}_{-0.04}$ & $2.83^{+0.06}_{-0.07}$ & $1.39^{+0.81}_{-0.85}$ & $42.44^{+0.03}_{-0.02}$ & $6.43^{+0.02}_{-0.03}$& $109^{+4}_{-4}$ &$0.66^{+0.02}_{-0.02}$&$0.75^{+0.03}_{-0.03}$&$0.81^{+0.02}_{-0.02}$&$0.20^{+0.08}_{-0.08}$ &  2012.41/1605 \\
& &  &  &  & &  & &  &  & \\
Su & $0.65^{+0.04}_{-0.05}$   & $1.65^{+0.03}_{-0.03}$ & $11.86^{+0.67}_{-0.77}$ & $43.91^{+0.05}_{-0.04}$ & $1.55^{+0.12}_{-0.13}$ & $1.37^{+0.32}_{-0.35}$ & $44.12^{+0.04}_{-0.05}$ & $6.39^{+0.05}_{-0.05}$& $56^{+3}_{-3}$ &$0.63^{+0.03}_{-0.03}$&$0.68^{+0.05}_{-0.05}$&--&--&  1502.18/1385 \\
& &  &  &  & &  & &  &  & \\
XMM2+Nu1 & $1.25^{+0.15}_{-0.16}$   & $1.73^{+0.02}_{-0.02}$ & $6.46^{+0.87}_{-0.89}$ & $43.10^{+0.02}_{-0.02}$ & $1.70^{+0.09}_{-0.10}$ & $6.15^{+0.16}_{-0.49}$ & $42.96^{+0.02}_{-0.03}$ & $6.43^{+0.03}_{-0.03}$& $142^{+9}_{-9}$ &$0.66^{+0.02}_{-0.02}$&$0.57^{+0.03}_{-0.03}$&$0.81^{+0.02}_{-0.02}$&$0.19^{+0.08}_{-0.08}$&  2734.39/2425 \\
& &  &  &  & &  & &  &  & \\
XMM3+Nu2 & $1.09^{+0.18}_{-0.19}$   & $1.76^{+0.02}_{-0.02}$ & $6.32^{+0.11}_{-0.07}$ & $43.09^{+0.02}_{-0.03}$ & $1.77^{+0.12}_{-0.14}$ & $5.28^{+0.46}_{-0.47}$ & $42.97^{+0.01}_{-0.01}$ & $6.44^{+0.03}_{-0.03}$& $164^{+5}_{-3}$ &$0.65^{+0.02}_{-0.02}$&$0.19^{+0.03}_{-0.03}$&$0.79^{+0.01}_{-0.01}$&$0.18^{+0.03}_{-0.03}$&  2394.72/2134 \\
& &  &  &  & &  & &  &  & \\
XRTc    & $2.28^{+0.15}_{-0.16}$   & $1.59^{+0.17}_{-0.06}$ & $5.49^{+0.31}_{-0.32}$ & $42.18^{+0.12}_{-0.13}$ & $1.60^{+0.21}_{-0.22}$ & $1.83^{+0.14}_{-0.38}$ & $42.38^{+0.08}_{-0.09}$ & $--$& $--$ &--&--&--&--&  176.68/189 \\
& &  &  &  & &  & &  &  & \\
XRT21    & $2.37^{+3.47}_{-0.91}$   & $1.62^{+0.21}_{-0.20}$ & $2.82^{+0.27}_{-0.28}$ & $42.87^{+0.08}_{-0.12}$ & $1.31^{+0.34}_{-0.33}$ & $1.08^{+0.24}_{-0.23}$ & $42.12^{+0.04}_{-0.04}$ & $--$& $--$ &--&--&--&--&  54.95/60 \\
& &  &  &  & &  & &  &  & \\
XRT24    & $2.26^{+0.77}_{-0.56}$   & $1.74^{+0.10}_{-0.18}$ & $2.82^{+0.62}_{-0.14}$ & $42.91^{+0.14}_{-0.15}$ & $1.37^{+0.37}_{-0.39}$ & $6.20^{+0.51}_{-0.54}$ & $42.57^{+0.04}_{-0.05}$ & $--$& $--$ &--&--&--&--&  43.18/74 \\
\hline
\end{tabular}
\begin{flushleft}
\footnotesize{$^\dagger$ in the unit of photons/keV/cm$^2$/s.}
\end{flushleft}
\end{table*}

After successfully fitting the observed spectra up to 10 keV, we extended the analysis into the high-energy domain (above 10 keV). In this range, we have three observations: a combined spectrum of XIS and HXD from the {\it Suzaku} observation (Su) in 2009, and two {\it NuSTAR} observations (Nu1 and Nu2) from 2015, conducted nearly simultaneously with the {\it XMM-Newton} observations (XMM2 and XMM3). These data are combined to study the high-energy spectral properties of Mrk~1040. When extrapolating the primary power-law model into the high-energy domain, we did not observe any significant deviation between the spectral data points and the model. The corresponding ratio plots between models are shown in the left panels of Figure~\ref{fig:chi}, where no notable deviation is found above 10 keV, in the Su, XMM2+Nu1, and XMM3+Nu2 observations. This suggests that the Compton hump or reflection component above $\sim 10$ keV \citep{Ross1993, Ross2005, Garcia2013, Garcia2014} is either absent or not significant during these observations.

After parameterizing the observed spectrum of Mrk~1040 at each epoch using simple phenomenological models such as \texttt{Powerlaw}, we applied more sophisticated physical models, including \texttt{TCAF} and \texttt{AGNSED}, to gain deeper insight into the underlying physical processes of the source. The results are discussed in the following sections.

\subsection{Power-law}
\label{sec: pl}
The \texttt{Powerlaw} model is a phenomenological approach used to characterize the spectral slope of the observed X-ray spectrum. Studies show that the photon index, $\Gamma$, for different types of AGNs typically lies in the range of $\Gamma \sim 1.5-2.5$ \citep{Nandra1994, Reeves2000, Piconcelli2005, Page2005}. This parameter is directly correlated with several physical properties of AGNs, such as the Eddington ratio \citep{Shemmer2006, Piconcelli2005, Shemmer2008, Porquet2004} and black hole mass \citep{Piconcelli2005, Porquet2004, Shemmer2006}. We began X-ray spectral fitting of Mrk~1040 across different epochs using an absorbed power-law model. The baseline model used for this fit is: \texttt{TBabs$\times$zTBabs$\times$Constant$\times$(Powerlaw+zGauss+Powerlaw)} and the corresponding results are summarized in Table~\ref{tab:powerlaw}. 

As shown in Figure~\ref{fig:chi}, a strong soft excess was detected in the 2009 XMM1 spectrum, consistent with the earlier detection reported by \citep{Tripathi2011}. We find distinct spectral slopes for the primary continuum (3.0–10.0 keV) and the soft excess (below 3 keV), with $\Gamma_{\rm PC} = 1.73 \pm 0.02$ and $\Gamma_{\rm WA} = 2.83^{+0.06}_{-0.07}$, respectively (Table~\ref{tab:powerlaw}). The steeper slope below 3 keV suggests the presence of excess emission in the soft X-ray band. While fitting the soft excess below 3 keV, we notice the prominent presence of edges at $0.66\pm0.02$ keV with an optical depth $0.75\pm0.03$ and at $0.81\pm0.02$ keV with an optical depth $0.20\pm0.08$, which are previously reported by \citet{Tripathi2011}. The luminosity of the primary continuum in the energy range of 3--10 keV ($L_{\rm PC}$) is estimated to be $2.63^{+0.34}_{-0.21} \times 10^{43}$ erg s$^{-1}$, while the luminosity of the soft excess component ($L_{\rm SE}$) is $2.82^{+0.25}_{-0.12} \times 10^{43}$ erg s$^{-1}$. The value of the intrinsic hydrogen column density is found to be $N_{\rm H} = 1.70^{+0.76}_{-0.72} \times 10^{21}$ cm$^{-2}$. A fluorescent Fe~K$\alpha$ line is detected at $6.43^{+0.03}_{-0.02}$ keV with an equivalent width of $109 \pm 4$ eV.

After a gap of 4 years, Mrk~1040 was observed by {\it Suzaku} in 2013 (Su). As no soft excess is detected in this observation (see Figure~\ref{fig:chi}), the photon indices for the primary continuum and the soft band are consistent within uncertainties: $\Gamma_{\rm PC} = 1.65 \pm 0.03$ and $\Gamma_{\rm WA} = 1.55\pm0.13$. The luminosities increased in both bands, with $L_{\rm PC} = 4.26^{+0.45}_{-0.42} \times 10^{45}$ erg s$^{-1}$ and $L_{\rm SE} = 5.88\pm0.45 \times 10^{44}$ erg s$^{-1}$. The intrinsic hydrogen column density is found to be $N_{\rm H} = 0.65^{+0.04}_{-0.05} \times 10^{21}$ cm$^{-2}$. An Fe~K$\alpha$ emission line is also detected at $6.39\pm0.05$ keV with an equivalent width of $56 \pm 3$ eV. During this observation, an edge at $0.63\pm0.03$ keV with an optical depth of $0.68\pm0.05$ is also detected, as in the case of the XMM1 observation. 

In the 2015 broadband observations (XMM2+Nu1 and XMM3+Nu2), we did not detect any statistically significant soft excess or reflection hump (see Figure~\ref{fig:chi}). This suggests either the soft excess component is mostly weak or largely masked by the absorption component. The power-law slopes across the 0.3--40 keV energy range remain consistent in both observations. We obtain $\Gamma_{\rm PC} = 1.73 \pm 0.02$ for XMM2+Nu1 and $\Gamma_{\rm PC} = 1.76 \pm 0.02$ for XMM3+Nu2. Due to the absence of a soft excess component, the soft band photon indices ($\Gamma_{\rm WA}$) are also comparable to the primary continuum slopes: $\Gamma_{\rm WA} = 1.7\pm 0.1$ for XMM2+Nu1 and $\Gamma_{\rm WA} = 1.77 \pm 0.14$ for XMM3+Nu2. While fitting the soft X-ray spectrum below 3 keV, we notice the presence of edges at $0.66\pm0.03$ keV with an optical depth of $0.57\pm0.03$ and at $0.81\pm0.01$ keV with an optical depth of $0.19\pm0.08$ in both observations, XMM2+Nu1 and XMM3+Nu2, as in the case of the XMM1 observation. The values of the intrinsic hydrogen column densities are found to be comparable: $N_{\rm H} = 1.25\pm0.16 \times 10^{21}$ cm$^{-2}$ and $N_{\rm H} = 1.09\pm0.19 \times 10^{21}$ cm$^{-2}$, respectively. An Fe~K$\alpha$ emission line is detected at $6.43 \pm 0.03$ keV and $6.44 \pm 0.03$ keV, with equivalent widths of $142 \pm 9$ eV and $164\pm4$ eV, respectively. The luminosity of the primary continuum is estimated as $L_{\rm PC} = 7.94\pm0.35 \times 10^{42}$ and $L_{\rm PC} = 3.98^{+0.35}_{-0.42} \times 10^{42}$ erg s$^{-1}$ with soft excess luminosity $L_{\rm SE} = 1.45^{+0.07}_{-0.09} \times 10^{43}$ and $L_{\rm SE} = 1.78\pm0.03 \times 10^{43}$ erg s$^{-1}$ for XMM2+Nu1 and XMM3+Nu2, respectively. 

In 2015, Mrk~1040 was also observed by {\it Swift}/XRT at multiple exposures, which are combined into a single observation labeled XRTc (see Section~\ref{sec:obs}). As no soft excess was detected in this observation, the power-law indices for the primary continuum and the soft band are found to be similar: $\Gamma_{\rm PC} = 1.56^{+0.17}_{-0.06}$ and $\Gamma_{\rm WA} = 1.60 \pm 0.22$. The estimated luminosities of the primary continuum and the soft excess band are $L_{\rm PC} = 1.51\pm0.38 \times 10^{42}$ erg s$^{-1}$ and $L_{\rm SE} = 2.40\pm0.12 \times 10^{43}$ erg s$^{-1}$, respectively. The intrinsic hydrogen column density for this combined observation is determined to be $N_{\rm H} = 2.28\pm0.16 \times 10^{21}$ cm$^{-2}$.

Later, {\it Swift}/XRT also observed Mrk~1040 in 2021 (XRT21) and 2024 (XRT24), each with only a few kiloseconds of exposure. Due to the short exposure time and limited spectral resolution, the Fe~K$\alpha$ line could not be detected in these observations. The photon indices of the primary continuum are found to be $\Gamma_{\rm PC} = 1.62 \pm 0.21$ for XRT21 and $\Gamma_{\rm PC} = 1.74 \pm 0.15$ for XRT24. The corresponding luminosities are $L_{\rm PC} = 1.20\pm0.09 \times 10^{43}$ erg s$^{-1}$ and $L_{\rm PC} = 8.13\pm0.81 \times 10^{42}$ erg s$^{-1}$, respectively. In the lower energy band, the power-law indices are $\Gamma_{\rm WA} = 1.31 \pm 0.34$ for XRT21 and $\Gamma_{\rm WA} = 1.37 \pm 0.39$ for XRT24, with corresponding luminosities of $3.02^{+0.35}_{-0.38} \times 10^{42}$ erg s$^{-1}$ and $3.71\pm0.32 \times 10^{42}$ erg s$^{-1}$. The power law model fitting parameters obtained from the spectral fitting are given in Table~\ref{tab:powerlaw}.

\subsection{Physical Models}
\label{sec:phymo}
From the phenomenological modeling of the observed X-ray spectra, we find that the power-law photon indices vary between $1.59^{+0.17}_{-0.06}$ and $1.76 \pm 0.02$, along with variations in luminosity across different energy bands. These variations suggest changes in the physical properties of the X-ray emitting region. To gain deeper insights into the physical characteristics of the X-ray emitting region, such as its size and geometry, we apply two physical models, \texttt{AGNSED} \citep{Kubota2018} and \texttt{TCAF} \citep{CT95}, to the observed spectra. These models allow us to probe the origin of different spectral components and the underlying causes of spectral variability over the observational period spanning 2009 to 2024.

\subsubsection{\texttt{AGNSED}}
\label{sec:agnsed}
The \texttt{AGNSED} model \citep{Kubota2018} is a physically motivated model designed to fit the X-ray spectra of AGNs. It provides a comprehensive framework to interpret the complex emission arising from the accretion flow and the associated Comptonizing regions. The model divides the accretion structure into three primary components: (i) an outer Novikov–Thorne standard disk \citep{Novikov1973}, which supplies the seed photons, (ii) an inner warm Comptonizing region responsible for producing the soft X-ray excess, and (iii) a hot corona that generates the hard X-ray power-law continuum. This model self-consistently incorporates both thermal and dynamical effects in the accretion flow to simulate the disk emission and Comptonization processes. During spectral fitting in \texttt{XSPEC}, the model allows several physical parameters to vary, including the temperatures of the warm and hot corona $(kT_{e,\mathrm{warm}}, kT_{e,\mathrm{hot}})$, the sizes of the warm and hot regions $(R_{\mathrm{warm}}, R_{\mathrm{hot}})$ in gravitational radius ($R_g$), and the mass accretion rate $\log \dot{m}$, defined as the logarithmic ratio of the actual accretion rate $(\dot{m})$ to the Eddington accretion rate $(\dot{m}_{\mathrm{Edd}})$. These parameters are constrained by the spectral indices $(\Gamma{\mathrm{warm}}, \Gamma_{\mathrm{hot}})$ of the corresponding Comptonizing regions. Additional intrinsic parameters include the black hole mass $(M_{\mathrm{BH}})$ in $M_\odot$, the black hole spin parameter $(a^*)$, the comoving distance in Mpc, the cosine of the inclination angle $(\cos i)$, and the logarithm of the outer disk radius $(\log r_{\mathrm{out}})$ in $R_g$. The model also includes the effect of reprocessing through a parameter called \texttt{reprocess}.

For this work, we adopt a black hole mass of $M_{\mathrm{BH}} = 4.5 \times 10^7M_\odot$ and a comoving distance of 71.1 Mpc, corresponding to a redshift of $z = 0.01667$ \citep{Huchra1999}. The inclination angle $\cos i$ and the outer disk radius are set to their default values, $\cos i = 0.5$ and $\log r_{\mathrm{out}} = -1$, respectively. Since no reflection component is present in the observed spectra (see Section~\ref{sec:characterization}), we set \texttt{reprocess} = 0. Finally, the model normalization is fixed to unity during spectral fitting.

The soft excess was only observed in the 2009 (XMM1) observation, indicating the presence of a warm corona in addition to the hot corona.  For this observation, we estimated the warm corona parameters as follows: temperature $kT_{\rm e,warm} = 0.26\pm0.2$ keV, spectral slope $\Gamma_{\rm warm} = 2.92^{+0.13}_{-0.05}$, and radius $R_{\rm warm} = 30.70^{+3.42}_{-5.09}~R_{\rm g}$. The corresponding parameters for the hot corona are $kT_{\rm e,hot} = 103.3^{+57.9}_{-51.4}$ keV, $\Gamma_{\rm hot} = 1.79 \pm 0.01$, and $R_{\rm hot} = 24.90^{+3.26}_{-4.72}~R_{\rm g}$. The normalized accretion rate is found to be $\log \dot{m} = -1.64^{+0.04}_{-0.05}$, and the black hole spin is estimated as $a = 0.67 \pm 0.05$. The detailed results are quoted in Table~\ref{tab:pm}. 

In the remaining observations, where no soft excess was detected, the warm corona parameters could not be constrained. As a result, $kT_{\rm e,warm}$, $\Gamma_{\rm warm}$, and $R_{\rm warm}$ were pegged at their lower limits during spectral fitting. To ensure stability in the fits, we froze these parameters at their lower bounds: $kT_{\rm e,warm} = 0.1$ keV, $\Gamma_{\rm warm} = 2.0$, and $R_{\rm warm} = 6.0~R_{\rm g}$, during the \texttt{AGNSED} model fitting of the \textit{Su}, \textit{XMM2+Nu1}, \textit{XMM3+Nu2}, \textit{XRTc}, \textit{XRT21}, and \textit{XRT24} observations (Table~\ref{tab:pm}).

During the 2013 {\it Suzaku} observation, the source luminosity is found to be higher compared to the previous epoch (see Table~\ref{tab:powerlaw}). Although a simple power-law model provides an acceptable fit to the observed spectrum with a reduced $\chi^2_\nu \approx 1$ (Table~\ref{tab:powerlaw}), fitting the data with the physical AGNSED model yielded a comparatively poorer fit with $\chi^2/\text{dof}=1851.3/1194$. Within the AGNSED framework, an increased accretion rate generally enhances the warm-corona component \citep{Hagen2024, Hagen2024a, Palit2024}. This implies that the optically thick inner disc extends down to small radii. The {\it Suzaku} spectrum instead favors a geometry with a truncated inner disc and an expanded hot Comptonizing region. Spectral fitting indicates that the normalized accretion rate increased from $\log \dot{m}=-1.64^{+0.04}_{-0.05}$ in 2009 to $-1.06^{+0.15}_{-0.13}$ in 2013, resulting in the enhanced X-ray luminosity (see Section~\ref{sec: pl}). To account for this higher luminosity, both the temperature and the size of the hot corona increased significantly, with $kT_{\rm e,hot}=152.3^{+37.5}_{-31.1}$ keV and $R{\rm hot}=46.9^{+4.8}_{-3.8}~R{\rm g}$, compared to the XMM1 observation. The larger corona implies a greater production of hard photons, consistent with the harder spectral index observed during this epoch, $\Gamma_{\rm hot}=1.61\pm0.01$ (see Table~\ref{tab:pm}).

We analyzed two broadband (XMM2+Nu1 and XMM3+Nu2) and a combined {\it Swift}/XRT observations (XRTc) observations in 2015. From the broadband fitting of XMM2+Nu1 and XMM3+Nu2 data, the hot corona temperatures are found to be $kT_{\rm e,hot} = 127.3^{+47.1}_{-42.6}$ and $125.6^{+16.2}_{-11.7}$ keV, with radii $R_{\rm hot} = 10.81^{+1.01}_{-1.91}$ and $12.92^{+1.48}_{-1.03}~R_{\rm g}$, respectively. The corresponding normalized accretion rates are estimated as $\log \dot{m} = -1.47^{+0.05}_{-0.09}$ and $-1.54\pm0.07$, respectively, both with $\Gamma_{\rm hot} = 1.61 \pm 0.01$. The MCMC analysis for the best fit {\tt AGNSED} model of the XMM3+NU2 observation is presented in Figure \ref{fig:mcmc_a}. For XRTc, we obtained $\log \dot{m} = -1.44^{+0.20}_{-0.34}$, $\Gamma{\rm hot} = 1.60^{+0.10}_{-0.08}$, $kT{\rm e,hot} = 129.1^{+76.8}_{-37.3}$ keV, and $R{\rm hot} = 12.65^{+9.21}_{-8.77}~R{\rm g}$. The Swift/XRT data during 2021–2024 show similar temperatures ($kT_{\rm e,hot} = 128–140$ keV) but larger radii ($R_{\rm hot} = 15–23$ $R_{\rm g}$) and modest changes in slope ($\Gamma_{\rm hot} = 1.6–1.7$) and accretion rate ($\log \dot{m} = -1.6$ to $-1.4$).

\begin{table*}
\setlength{\tabcolsep}{3.0pt}
 \renewcommand{\arraystretch}{2.0}
\begin{center}
\caption{Best fit parameters for observations of Mrk~1040 with the physical models. The baseline model we used to fit the observed spectra is: \texttt{TBabs$\times$zTBabs$\times$Constant$\times$(AGNSED+zGauss)$\times$Edge1$\times$Edge2}}
\label{tab:pm}
\begin{tabular}{llccccccc}
\hline
Model & Parameter & XMM1 & Su & XMM2+Nu1 & XMM3+Nu2 & XRTc & XRT21 & XRT24 \\
\hline
\texttt zTbabs & ${\rm N}_{\rm H}~(10^{21}\text{cm}^{-2})$ & $1.71^{+0.74}_{-0.75}$ & $0.84^{+0.07}_{-0.03}$ & $1.29^{+0.19}_{-0.26}$ & $1.01^{+0.20}_{-0.22}$ & $2.22^{+0.29}_{-0.30}$ & $1.99^{+3.55}_{-1.74}$ & $1.98^{+0.89}_{-0.87}$ \\
\texttt{AGNSED} & $\log \dot{m}$ & $-1.64^{+0.04}_{-0.05}$ & $-1.06^{+0.15}_{-0.13}$ & $-1.47^{+0.05}_{-0.09}$ & $-1.54^{+0.07}_{-0.07}$ & $-1.44^{+0.20}_{-0.34}$ & $-1.61^{+0.57}_{-0.59}$ & $-1.39^{+0.18}_{-0.27}$ \\
& $a$ & $0.69^{+0.05}_{-0.18}$ & $0.65^{+0.20}_{-0.28}$ & $0.81^{+0.10}_{-0.17}$ & $0.75^{+0.06}_{-0.08}$ & $0.75^f$ & $0.75^f$ & $0.75^f$ \\
& $KT_\text{e,hot}~\text{(keV)}$ & $103.3^{+57.9}_{-51.4}$ & $152.32^{+37.5}_{-41.1}$ & $127.3^{+47.1}_{-42.6}$ & $135.0^{+46.8}_{-24.0}$ & $129.1^{+76.8}_{-37.3}$ & $128.3^{+64.5}_{-26.4}$ & $139.7^{+81.9}_{-54.7}$ \\
& $KT_\text{e,warm}~\text{(keV)}$ & $0.26^{+0.2}_{-0.1}$ & $0.1^f$ & $0.1^f$ & $0.1^f$ & $0.1^f$ & $0.1^f$ & $0.1^f$ \\
& $\Gamma_\text{hot}$ & $1.79^{+0.01}_{-0.01}$ & $1.61^{+0.01}_{-0.01}$ & $1.73^{+0.01}_{-0.01}$ & $1.78^{+0.01}_{-0.01}$ & $1.60^{+0.10}_{-0.08}$ & $1.71^{+0.11}_{-0.10}$ & $1.60^{+0.15}_{-0.13}$ \\
& $\Gamma_\text{warm}$ & $2.92^{+0.13}_{-0.05}$ & $2.0^f$ & $2.0^f$ & $2.0^f$ & $2.0^f$ & $2.0^f$ & $2.0^f$  \\
& $\text{R}_\text{Hot}~(R_g)$ & $24.90^{+3.26}_{-4.72}$ & $46.38^{+4.84}_{-3.83}$ & $10.81^{+1.01}_{-1.91}$ & $12.60^{+1.48}_{-1.03}$ & $12.65^{+9.21}_{-8.77}$ & $22.74^{+16.51}_{-10.82}$ & $14.97^{+09.21}_{-10.46}$ \\
& $\text{R}_\text{warm}~(R_g)$ & $30.70^{+3.42}_{-5.09}$ & $6.0^f$ & $6.0^f$ & $6.0^f$ & $6.0^f$ & $6.0^f$ & $6.0^f$ \\
\texttt {Edge1}&$E_{1} (keV)$&$0.66^{+0.02}_{-0.02}$&$0.63^{+0.01}_{-0.01}$&$0.63^{+0.01}_{-0.01}$&$0.64^{+0.05}_{-0.05}$&--&--&--  \\
&$\tau_{1}$   &$0.75^{+0.03}_{-0.03}$&$0.68^{+0.07}_{-0.04}$&$0.57^{+0.03}_{-0.02}$&$0.54^{+0.03}_{-0.03}$&--&--&--\\
\texttt {Edge2}&$E_{2} (keV)$&$0.81^{+0.03}_{-0.03}$&--&$0.81^{+0.01}_{-0.01}$&$0.79^{+0.01}_{-0.01}$&--&--&--  \\
&$\tau_{2}$   &$0.20^{+0.02}_{-0.02}$&--&$0.19^{+0.03}_{-0.02}$&$0.18^{+0.03}_{-0.02}$&--&--&--\\
\hline
& $\chi^2/\text{dof}$ & $2041.61/1618$ & $1851.32/1194$ & $2718.09/2482$ & $2374.66/2105$ & $169.99/186$ & $51.43/64$ & $48.67/78$ \\
\hline
\hline
\texttt{zTbabs} & ${\rm N}_{\rm H}~(10^{21}\text{cm}^{-2})$ & $1.70^{+0.77}_{-0.78}$ & $0.86^{+0.05}_{-0.09}$ & $1.33^{+0.21}_{-0.25}$ & $0.97^{+0.52}_{-0.54}$ & $2.84^{+0.49}_{-0.44}$ & $1.97^{+2.59}_{-1.47}$ & $1.98^{+0.74}_{-0.98}$ \\
\texttt{TCAF} & $M_\text{BH}~(\times10^{7}\dot{M}_\odot)$ & $4.51^{+2.81}_{-2.61}$ & $4.27^{+1.12}_{-1.49}$ & $5.02^{+1.01}_{-1.03}$ & $4.48^{+1.08}_{-1.07}$ & $4.21^{+3.54}_{-3.25}$ & $4.19^{+5.38}_{-5.55}$ & $4.55^{+3.48}_{-3.36}$ \\
& $\dot{m}_d~(\times10^{-2}\dot{m}_\text{Edd})$ & $1.19^{+0.11}_{-0.10}$ & $5.62^{+0.45}_{-0.49}$ & $0.91^{+0.02}_{-0.03}$ & $0.97^{+0.03}_{-0.02}$ & $1.20^{+0.25}_{-0.21}$ & $1.30^{+0.53}_{-0.52}$ & $1.81^{+0.31}_{-0.36}$ \\
& $\dot{m}_h~(\times10^{-2}\dot{m}_\text{Edd})$ & $3.10^{+0.28}_{-0.29}$ & $13.9^{+0.18}_{-0.12}$ & $1.79^{+0.21}_{-0.20}$ & $1.76^{+0.18}_{-0.23}$ & $2.15^{+0.45}_{-0.52}$ & $3.39^{+0.68}_{-0.65}$ & $3.44^{+0.51}_{-0.54}$ \\
& $X_s~(R_g)$ & $24.07^{+2.20}_{-2.21}$ & $42.51^{+2.78}_{-2.82}$ & $12.63^{+1.07}_{-2.58}$ & $11.42^{+1.89}_{-1.92}$ & $11.28^{+5.45}_{-2.61}$ & $23.01^{+8.47}_{-9.87}$ & $19.82^{+4.52}_{-3.21}$ \\
& $R$ & $2.20^{+0.93}_{-1.04}$ & $2.21^{+0.78}_{-0.82}$ & $2.22^{+0.62}_{-0.15}$ & $2.29^{+0.67}_{-0.23}$ & $2.15^{+0.53}_{-0.36}$ & $2.11^{+0.87}_{-0.84}$ & $2.07^{+0.61}_{-0.66}$ 
\\
& $\text{Norm}^\dagger~(\times10^{-5})$ & $5.01^{+0.33}_{-0.52}$ & $3.61^{+0.21}_{-0.25}$ & $3.56^{+0.05}_{-0.06}$ & $3.39^{+0.41}_{-0.46}$ & $4.42^{+0.10}_{-0.09}$ & $5.08^{+0.15}_{-0.02}$ & $3.82^{+0.21}_{-0.27}$ \\
\texttt {Edge1}&$E_{1} (keV)$&$0.66^{+0.02}_{-0.02}$&$0.63^{+0.01}_{-0.01}$&$0.63^{+0.01}_{-0.01}$&$0.64^{+0.01}_{-0.01}$&--&--&--  \\
&$\tau_{1}$   &$0.75^{+0.03}_{-0.03}$&$0.68^{+0.07}_{-0.04}$&$0.57^{+0.03}_{-0.02}$&$0.53^{+0.03}_{-0.03}$&--&--&--\\
\texttt {Edge2}&$E_{2} (keV)$&$0.81^{+0.03}_{-0.03}$&--&$0.81^{+0.01}_{-0.01}$&$0.77^{+0.01}_{-0.01}$&--&--&--  \\
&$\tau_{2}$   &$0.20^{+0.02}_{-0.02}$&--&$0.21^{+0.03}_{-0.02}$&$0.24^{+0.03}_{-0.02}$&--&--&--\\
\hline
& $\chi^2/\text{dof}$ & $2051.82/1608$ & $1451.32/1194$ & $2724.70/2466$ & $2369.74/2106$ & $169.99/186$ & $50.16/57$ & $44.26/73$ \\
\hline
\end{tabular}
\end{center}
\flushleft \textbf{Note:} $^\dagger$ in the unit of photons/keV/cm$^2$/s. $f$ indicates a frozen parameter.
\end{table*}

\subsection{\texttt{TCAF}}
\label{sec:tcaf}
During the \texttt{AGNSED} model fitting of the observed spectra, we found that the variations in luminosity and spectral slope depend on the accretion flow parameters and the physical size of the corona. Depending on viscosity, the accretion flow may segregate into two components: a portion of the inflowing matter with higher viscosity settles in the equatorial plane, forming a geometrically thin, optically thick Keplerian disc, while the remaining matter, possessing relatively low angular momentum, forms a sub-Keplerian flow \citep{Chakrabarti1989, Chakrabarti1990, CT95}. This theoretical possibility is realized in the Two Component Advective Flow (TCAF) model \citep{CT95}. The TCAF is a physically motivated accretion paradigm used for fitting X-ray spectra of black holes across a wide range of masses, from stellar-mass black holes in X-ray binaries \citep{Debnath2014, Debnath2015, Mondal2014, Mondal2016, Chatterjee2016, Jana2016, Molla2016}, to supermassive black holes in AGNs \citep{Mandal2008, Nandi2019, Nandi2021, Nandi2024}. Based on the transonic flow theory \citep{Chakrabarti1989, Chakrabarti1990}, the model incorporates both hydrodynamical and radiative processes in a self-consistent manner. It consists of a Keplerian disc \citep{SS73} surrounded by a sub-Keplerian, low angular momentum halo \citep{CT95}. The Keplerian disc is truncated at the centrifugal barrier, where a shock may form. The post-shock region, referred to as the CENtrifugal pressure-supported BOundary Layer (CENBOL), acts as the Comptonizing region (i.e., the corona), which up-scatters soft photons from the disc to produce hard X-ray emission.

The TCAF model self-consistently explains spectral states and variability by varying four key physical parameters: the disk accretion rate $(\dot{m}_d)$ and the sub-Keplerian halo accretion rate $(\dot{m}_h)$ (both in units of the Eddington accretion rate, $\dot{m}_{\mathrm{Edd}}$), the shock location $(X_s)$ in gravitational radii ($R_g$), and the compression ratio $(R)$, which is the ratio of post-shock to pre-shock flow densities. The model also depends on intrinsic parameters such as the mass of the central black hole $(M_{\mathrm{BH}})$ in solar masses ($M_\odot$), and a normalization parameter that scales the observed spectrum to the theoretical TCAF spectrum. The upper and lower limits, along with the default values of each parameter, are stored in a file named \texttt{lmodel.dat}, as summarized in Table~\ref{tab:tcaf_par}. To read this table into \texttt{XSPEC}, we execute the commands \texttt{initpackage} and \texttt{lmod} to load the model and fit the observed spectra of Mrk~1040. The corresponding best-fitting spectral results are presented in Table~\ref{tab:pm}. To determine the optimal spectral parameters, we ran the source code approximately $10^5$ times using the $\chi^2$ minimization technique.

\begin{table}
\centering
\caption{The TCAF parameter space as defined in the file \texttt{lmod.dat}. Two bounds are provided for each parameter: soft and hard limits used during the iteration process.}
\label{tab:tcaf_par}
\begin{tabular}{@{}lccccc@{}}
\hline
Parameter & Units & Default & Min & Max & Increment \\
\hline
$M_{\rm BH}$     & $M_\odot$     & $1.0 \times 10^7$ & $1.0 \times 10^4$ & $9.9 \times 10^9$ & 10.0 \\
$\dot{m}_{\rm d}$ & $\dot{m}_{\rm Edd}$ & 0.001 & 0.0001 & 5.0 & 0.0001 \\
$\dot{m}_{\rm h}$ & $\dot{m}_{\rm Edd}$ & 0.01 & 0.0001 & 5.0 & 0.0001 \\
$X_s$            & $R_{\rm g}$   & 50.0 & 8.0 & 1000.0 & 0.5 \\
$R$              & --            & 2.5  & 1.01 & 6.8 & 0.1 \\
\hline
\end{tabular}
\end{table}

\begin{figure*}
\begin{center}
\includegraphics[trim={0 1.8cm 0cm 0},scale=1.3]{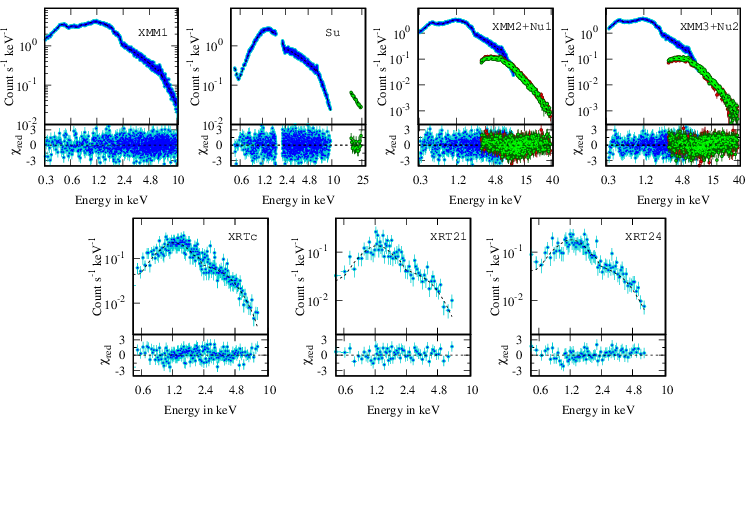}
\caption{\texttt{TCAF} model fitted spectra of Mrk 1040 from the different epochs of observations along with the residuals. }
\label{fig:spec}
\end{center}
\end{figure*}

We began the spectral fitting of Mrk~1040 using the \texttt{TCAF} model on the 2009 (XMM1) observation. For this observation, the disc and halo accretion rates are found to be $\dot{m}_d = 1.19^{+0.11}_{-0.10} \times 10^{-2}~\dot{m}_{\rm Edd}$ and $\dot{m}_h = 3.10^{+0.28}_{-0.29} \times 10^{-2}~\dot{m}_{\rm Edd}$, respectively. The Compton cloud or corona is found to extend up to $X_s = 24.04^{+2.20}_{-2.21}~r_g$, with a compression ratio of $R = 2.20^{+0.93}_{-1.04}$. The values of each parameter are presented in Table~\ref{tab:pm}, and the corresponding model fitted spectrum with the residual is presented in Figure~\ref{fig:spec}.  

Next, we analyzed a broadband observation (Su), covering the energy range of 0.5--30 keV. In this observation, we detected a sudden increase in the accretion rates for both components of the flow. The disc accretion rate increased from $1.19^{+0.11}_{-0.10} \times 10^{-2}$ to $5.62^{+0.45}_{-0.49} \times 10^{-2}~\dot{m}_{\rm Edd}$, while the halo accretion rate increased from $3.10^{+0.28}_{-0.29} \times 10^{-2}$ to $13.90^{+0.18}_{-0.12} \times 10^{-2}~\dot{m}_{\rm Edd}$. This enhancement in accretion flow led to a more prominent Compton cloud or corona. Consequently, the Compton cloud was observed to extend up to $X_s = 42.51^{+2.78}_{-2.82}~r_g$, significantly larger than in the previous observation. As a consequence, an increase in X-ray luminosity is also recorded, as noted in Table~\ref{tab:powerlaw}. The compression ratio for this observation is estimated to be $R = 2.21^{+0.78}_{-0.82}$, with the model normalization of $\text{Norm} = 3.61^{+0.21}_{-0.25} \times 10^{-5}$ photons keV$^{-1}$ cm$^{-2}$ s$^{-1}$.

In 2015, following a two-year gap, Mrk~1040 was observed nearly simultaneously by \textit{XMM-Newton} and \textit{NuSTAR}. We used data from these observations for a broadband spectral analysis covering the 0.2--40.0 keV energy range. During spectral fitting using the \texttt{TCAF} model, we found that the spectral parameters do not vary (or vary within the uncertainties) between the two observations. The disc accretion rates are estimated to be $0.91^{+0.02}_{-0.03} \times 10^{-2}$ and $0.97^{+0.03}_{-0.02} \times 10^{-2}~\dot{m}_{\rm Edd}$ for XMM2+Nu1 and XMM3+Nu2, respectively. The halo accretion rates are $1.76^{+0.21}_{-0.20} \times 10^{-2}$ and $1.94^{+0.18}_{-0.23} \times 10^{-2}~\dot{m}_{\rm Edd}$ for the same epochs. The shock location, which represents the outer boundary of the Compton cloud, remains nearly the same (within uncertainties), with $X_s = 12.63^{+1.07}_{-2.58}~r_g$ for XMM2+Nu1 and $X_s = 11.42^{+1.89}_{-1.92}~r_g$ for XMM3+Nu2. The compression ratios are found to be $R = 2.22^{+0.67}_{-0.23}$ and $R = 2.29^{+0.53}_{-0.36}$ for the respective observations. The values of each parameter are presented in Table~\ref{tab:pm}, and the corresponding model fitted spectrum with the residual is presented in Figure~\ref{fig:spec}. The MCMC analysis of the XMM3+NU2 observation for the best fit {\tt TCAF} model is presented in Figure \ref{fig:mcmc}.

In the same year, 2015, the {\it Swift}/XRT also observed Mrk~1040 at multiple epochs. We used the combined spectra to study the broadband spectral properties of the source. From spectral fitting using the \texttt{TCAF} model, the disc and halo accretion rates are found to be $\dot{m}_d = 1.20^{+0.25}_{-0.21} \times 10^{-2}\dot{m}_{\rm Edd}$ and $\dot{m}_h = 2.15^{+0.45}_{-0.52} \times 10^{-2}~\dot{m}_{\rm Edd}$, respectively. The size of the Compton cloud is estimated to be $X_s = 11.28^{+5.45}_{-2.61}~r_g$, with the compression ratio of $R = 2.15^{+0.53}_{-0.36}$. The detailed values of all parameters are presented in Table~\ref{tab:pm}.

The {\it Swift}/XRT also observed Mrk~1040 in 2021 (XRT21) and 2024 (XRT24). We used these observations to investigate the accretion geometry using the \texttt{TCAF} model. In these epochs, we found a steady increase in the disc accretion rate, while the halo rate remained nearly constant within uncertainties. The disc accretion rates are estimated to be $1.30^{+0.53}_{-0.52} \times 10^{-2}$ and $1.81^{+0.31}_{-0.36} \times 10^{-2}~\dot{m}_{\rm Edd}$, and the halo accretion rates are $3.39^{+0.68}_{-0.65} \times 10^{-2}$ and $3.44^{+0.51}_{-0.54} \times 10^{-2}~\dot{m}_{\rm Edd}$ for XRT21 and XRT24, respectively. With the increase in accretion rates compared to previous observations, we also observe that the shock location, representing the outer boundary of the Compton cloud, appears at $23.01^{+8.47}_{-9.87}~r_g$ and $19.82^{+4.52}_{-3.21}~r_g$ for XRT21 and XRT24, respectively. However, the compression ratio remains nearly the same in both cases, with $R = 2.11^{+0.87}_{-0.84}$ and $R = 2.07^{+0.21}_{-0.27}$. This suggests that despite the increased accretion rates, the corresponding enhancement in soft photons from the disc is insufficient to significantly cool the Compton cloud. As a result, the Compton cloud remains extended during these observations. The values of each parameter are presented in Table~\ref{tab:pm}, and the corresponding model fitted spectrum with the residual is presented in Figure~\ref{fig:spec}.

\begin{figure}
\begin{center}
\hspace*{-1.0 cm}
\includegraphics[trim={0 1cm 0cm 0},scale=1.1]{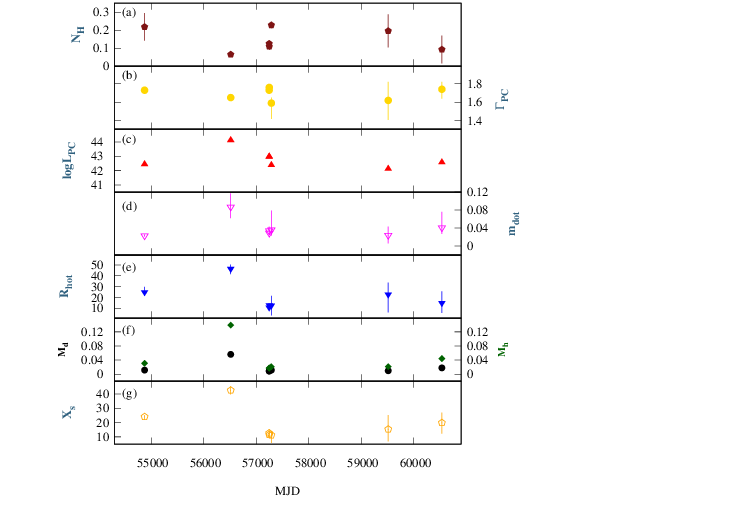}   
\caption{Variation of different spectral fitting parameters with time. The corresponding values of each parameter are given in Table~\ref{tab:pm}.}
\label{fig:multiplot}
\end{center}
\end{figure}

\section{Discussion}
\label{sec:discussion}
Mrk~1040 remains a relatively unexplored AGN, with only a few previous studies conducted, including the detection of soft excess in a 2009 observation \citep{Tripathi2011}. However, no detailed follow-up investigations have been conducted since then. In this work, we explore the high-energy X-ray regime to probe the physical processes occurring around the central supermassive black hole in Mrk~1040. Furthermore, we find a significant variation in the accretion rate over the 15-year observation period (2009--2024). This long-term monitoring reveals systematic changes in several physical parameters derived from both spectral and temporal analyses. In the following section, we discuss the physical interpretation of these variations found in this work.

\subsection{Evolution of the source with time}
\label{sec:evolution}
In this work, we analyze multi-epoch X-ray observations of Mrk~1040 over a 15-year duration (2009–2024) using \textit{XMM-Newton}, \textit{Suzaku}, \textit{NuSTAR}, and \textit{Swift} observatories. Our analysis reveals complex and evolving spectral and temporal characteristics of the source, primarily driven by variations in the accretion rate and coronal geometry. Our analysis reveals that the soft excess component, which was prominent in 2009, disappeared in subsequent observations. The highest X-ray luminosity was recorded during the 2013 (Su) observation. Broadband observations in 2015 revealed no evidence of a reflection component, although absorption features and a persistent Fe~K$\alpha$ line are still observed. The absence of these features in later observations may be due to the low exposure time and limited spectral resolution of the {\it Swift}/XRT instrument. In general, our results suggest that the physical properties of Mrk~1040 have undergone significant changes over the observation period.

In the 2009 (XMM1) observation, we observed a prominent soft excess component below 3 keV, consistent with the finding of \citet{Tripathi2011}. Our spectral modeling (see Section \ref{sec:spec} and Table \ref{tab:pm}) revealed two distinct photon indices for the soft excess and the primary continuum, indicating that these components have different emission mechanisms. We observed absorption edges at 0.66 keV and 0.81 keV, along with a narrow Fe~K$\alpha$ line at 6.43 keV, suggesting the presence of ionized material and/or reflection features during this epoch. However, the limited high-energy coverage (above 10 keV) makes it difficult to draw definitive conclusions about the reflection properties. Using both phenomenological and physical models (Sections~\ref{sec: pl} \& \ref{sec:phymo}), we identified a warm Comptonizing corona responsible for the soft excess emission and a hotter inner corona generating the hard X-ray continuum. The accretion-rate estimates indicate a dominance of sub-Keplerian flow ($\dot{m}_{\rm h}>\dot{m}_{\rm d}$) with a moderately spinning black hole (see Table \ref{tab:pm}). The timing analysis (Section~\ref{sec:time}) supports this geometry. A moderate zero-lag correlation between the soft band ($<3$ keV) and the primary continuum is consistent with a two-corona structure. The correlation strength decreases with increasing energy separation in the primary continuum, implying a layered Comptonizing region, possibly due to gradients in the electron temperature or optical depth. These results are consistent with a multi-zone coronal geometry \citep{Liu2011,Petrucci2018,Garcia2019,Nandi2021}, where the soft excess arises from reprocessing of disc photons in a warm, extended outer corona, and the harder photons originate from the compact, hot inner region near the black hole.

The 2013 ({\it Suzaku}) observation captures Mrk~1040 in a distinctly brighter X-ray phase than in 2009, coinciding with the disappearance of the previously detected soft-excess component. Spectral fitting (see Section~\ref{sec:spec} and Table~\ref{tab:powerlaw}) shows consistent photon indices in both the soft and hard bands, indicating that the emission likely arises from a single and hot Comptonizing region. Under the {\tt AGNSED} framework, the model provides a comparatively poor fit, whereas the {\tt TCAF} model yields an acceptable fit (see Table~\ref{tab:pm}), confirming that the observed spectrum is better described by a geometry dominated by a hot, optically thin Comptonizing region rather than by a warm-corona configuration assumed in {\tt AGNSED}. The enhanced luminosity is attributed to higher disc and halo accretion rates, which produce a hotter and more extended corona. The resulting hard X-ray continuum and expansion of the Comptonizing region are consistent with a truncated accretion disc, where the inner disc does not reach the innermost stable circular orbit. This configuration suppresses strong relativistic reflection from the innermost regions. Consequently, only a narrow Fe~K$\alpha$ line is detected, consistent with reflection from distant, colder material such as the outer disc \citep{George1991, Nandra2006} or the BLR \citep{Yaqoob2004, Liu2010}. The presence of this line, along with a soft X-ray absorption edge, further indicates ionized material in the vicinity of the X-ray source (see Table~\ref{tab:nh}). These results support a disc–corona geometry in which a vertically extended, hot Comptonizing region dominates the inner accretion flow, reducing the efficiency of reflection from the innermost disc. Overall, the 2013 X-ray observation supports a scenario in which enhanced accretion drives the formation of a thermally dominant and geometrically extended corona that explains the observed spectral behavior.

In this work, we analyzed two broadband observations ({\it XMM2+Nu1} and {\it XMM3+Nu2}) and a combined {\it Swift} dataset (XRTc) from 2015. Across all observations, the spectra showed no signature of a soft excess or reflection hump. The soft and hard X-ray photon indices are nearly identical (see Section~\ref{sec: pl} and Table~\ref{tab:powerlaw}), implying that both components arise from the same physical origin, most likely a compact, hot corona. During this epoch, a narrow Fe~K$\alpha$ line was detected along with soft X-ray absorption edges, suggesting the presence of ionized material along the line of sight. Physical modeling (Section~\ref{sec:phymo}) confirms a hot and compact Comptonizing region with low disc and halo accretion rates (Table~\ref{tab:pm}). The TCAF fits indicate a stable Compton cloud configuration bounded by a centrifugal pressure–supported shock close to the black hole, consistent with a geometrically compact inner flow.

Although {\it XMM1} and {\it XMM2+Nu1} show comparable luminosities, the \texttt{AGNSED}-derived accretion rates differ marginally as the single normalized accretion rate parameter in this model gets adjusted to the changes in the coronal geometry and radiative efficiency. The \texttt{TCAF} model, on the other hand, separately tracks the Keplerian and sub-Keplerian inflows and therefore does not map one-to-one onto the \texttt{AGNSED} model. In the 2015 epoch, the warm-corona parameters in the \texttt{AGNSED} model were unconstrained and frozen, leading to modest compensation through $\log \dot{m}$. This small shift is well within the systematic modeling uncertainties and does not alter the inferred transition from a warm-corona dominated to a compact, thermally stable corona.

Temporal analysis of the broadband data revealed low-amplitude variability across the entire energy range (see Section~\ref{sec:fvar}). Variations are slightly more pronounced below 8 keV, whereas the {\it NuSTAR} data showed no significant variability. The strong zero-lag cross-correlation between the soft (0.3--3 keV) and primary continuum (3--10 keV) bands indicates a tightly coupled emission region at low energies, while the absence of correlation above 10 keV suggests that the higher energy photons beyond 10 keV originate from a more stable inner corona. Collectively, these results indicate that the 2015 epoch corresponds to a quiescent phase in the accretion history of Mrk~1040. The lack of soft excess and reflection features, together with the reduced variability and consistent spectral slopes across energy bands, supports a truncated-disc geometry in which a compact, thermally stable Compton cloud dominates the X-ray emission under a sub-Eddington accretion regime.

In 2021 and 2024, Mrk~1040 was observed with {\it Swift}/XRT (XRT21 and XRT24, see Table~1). Owing to short exposures and limited spectral resolution, no Fe~K$\alpha$ emission line was detected in either epoch. Spectral fitting (see Section~\ref{sec: pl} and Table~\ref{tab:powerlaw}) indicates comparable photon indices in the soft and hard bands, with moderate luminosities. Although no prominent spectral features are identified, physical modeling (Section~\ref{sec:phymo}) reveals a hot and extended corona consistent with an enlarged Comptonizing region during these epochs.

The accretion flow analysis (Table~\ref{tab:pm}) shows that the halo accretion rate remained nearly steady, while the disc rate increased modestly from 2021 to 2024. This gradual enhancement of disc flow likely increased the soft-photon flux, producing a marginal steepening of the X-ray continuum and a slight rise in soft X-ray luminosity. The Compton cloud remained extended and quasi-stable, with the shock structure maintaining its compression ratio across both observations. These results indicate that Mrk~1040 was moderately active during these observations. The mild increase in disc accretion suggests a progressive refilling of the inner disc following a previous truncation. The lack of short-term variability and the absence of reflection or soft-excess signatures imply a radiatively inefficient, thermally stable corona dominating the X-ray output, consistent with a sub-Eddington accretion regime.

\subsection{Evolution of the Fe K$\alpha$ and Absorption Edges}
\label{sec:fe}

Over a span of 15 years (2009--2024), multi-epoch X-ray observations of Mrk~1040 reveal a gradually evolving Fe~K$\alpha$ emission line profile that reflects long-term changes in the accretion geometry and coronal structure. In the 2009 {\it XMM-Newton} observation, the Fe~K$\alpha$ line was detected at 6.4 keV with moderate equivalent width, accompanied by a strong soft excess and prominent ionized absorption edges — signatures of significant reprocessing within the inner disc region (see Section~\ref{sec: pl} and Table~\ref{tab:powerlaw}). In 2013, during the {\it Suzaku} epoch, the line weakened considerably (Table~\ref{tab:pm}), coinciding with the disappearance of the soft excess and the emergence of a hotter and more extended corona (see Section~\ref{sec:phymo}). The spectral flattening and increased coronal size suggest an expanded Comptonizing region that likely truncated the inner disc. This truncation reduces the efficiency of reflection, leading to a weaker Fe~K$\alpha$ line \citep{Markowitz2009, Lobban2010, Garcia2015, Xu2020}. However, detection of this line indicates that reflection still occurs from distant material, most likely the outer accretion disc or other cold reprocessing regions \citep{George1991, Matt1991}. The 2013 spectrum also shows the near-disappearance of the absorption edges that are seen in 2009, indicating a reduction in the opacity of the partially ionized absorber.

\begin{figure}
\begin{center}
\includegraphics[trim={0 2.4cm 0cm 0},scale=0.9]{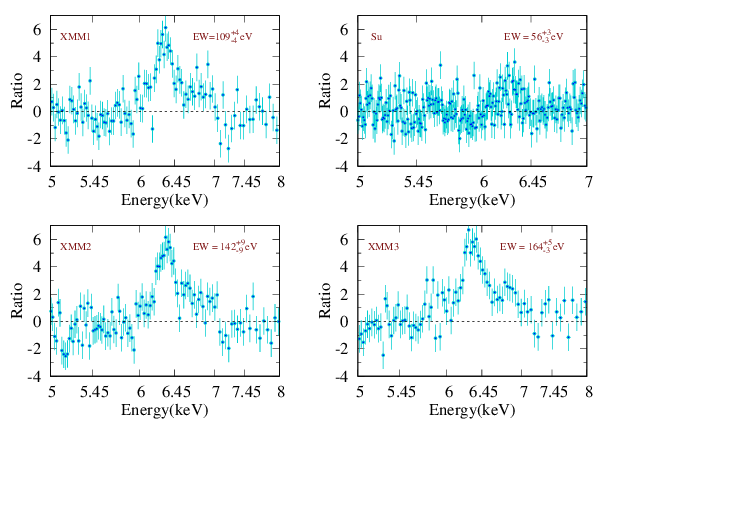}
\caption{Variation of Fe~K$\alpha$ emission line with time for Mrk~1040. The corresponding values of each parameter are given in Table~\ref{tab:powerlaw}.}
\label{fig:Fe}
\end{center}
\end{figure}

In the 2015 broadband {\it XMM-Newton} + {\it NuSTAR} observations, the Fe~K$\alpha$ line reappeared with moderate strength despite the absence of the soft excess and reflection hump. Compared to 2009, the line appeared in the 2015 broadband spectra is broader and exhibits a slightly larger equivalent width (Table~\ref{tab:pm}), consistent with the fluorescence from a more extended region of the disc once the strong absorption edges have weakened. The compact coronal structure and stable accretion rates inferred from the physical modeling (Section~\ref{sec:phymo}, Table~\ref{tab:pm}) suggest a quasi-steady geometry in which the line originates from reprocessing in distant, colder material, such as the outer disc or the broad-line region \citep{George1991, Matt1991, Yaqoob2004}. The absence of absorption edges in the 2015 spectra further supports a cleaner line of sight, allowing the broader Fe~K$\alpha$ component to dominate. During the 2021--2024 {\it Swift}/XRT observations, the Fe~K$\alpha$ line was not detected, likely due to the short exposures and limited spectral resolution of the instrument. The long-term evolution of the Fe~K$\alpha$ line (Figure~\ref{fig:Fe}) is therefore consistent with a gradual transition from radiatively efficient to inefficient accretion states, reflecting a change in the inner flow configuration from a complex and multi-zone corona to a thermally stable and compact Comptonizing region. This unified picture connects the changing Fe~K$\alpha$ width, equivalent width, and absorption-edge behaviour to the evolving disc–corona–absorber geometry across epochs.

\subsection{Evolution of ${\rm N}_{\rm H}$ }
\label{sec:nh}
We investigated the properties of the intrinsic absorber in Mrk~1040 over this long-term observational period. Our spectral analysis reveals significant temporal variation in the intrinsic hydrogen column density ($N_{\rm H}$). The measured $N_{\rm H}$ ranges from $0.65^{+0.07}_{-0.03} \times 10^{21}~\text{cm}^{-2}$ in the 2013 observation to $2.37^{+3.35}_{-1.75} \times 10^{21}\text{cm}^{-2}$ during the XRT21 observation in 2021. These values are initially obtained using the \texttt{zTbabs} model during spectral fitting (see Section~\ref{sec:characterization}). To further examine the absorber properties, we replaced the \texttt{zTbabs} model with more physically motivated models, \texttt{pcfabs} and \texttt{zxipcf}. The \texttt{pcfabs} model includes a covering fraction parameter ($C_f$), which helps to assess whether the X-ray emitting region is fully or partially covered, while the \texttt{zxipcf} model allows us to simultaneously constrain ionization, covering fraction, and $N_{\rm H}$. We tested these alternative models in place of \texttt{zTbabs} across all observations and spectral models. However, we found that neither the fit statistics improved nor the best-fit parameters changed significantly. Therefore, the results presented in Table~\ref{tab:nh} are based on fits using the simpler power-law models. 

From the spectral analysis, we found that $N_{\rm H}$ was the lowest during the 2013 observation. This epoch corresponds to a high-luminosity state, likely associated with an expanded hot corona. The reduced absorption observed at that time may be due to a geometrically thinned or displaced absorber, possibly caused by the coronal expansion, which displaced the surrounding material. However, later observations in 2021 (XRT21) and 2024 (XRT24) showed higher values of $N_{\rm H}$, measured at $2.37^{+3.35}_{-1.75} \times 10^{21}$ and $2.26^{+0.98}_{-0.87} \times 10^{21}~\text{cm}^{-2}$, respectively, suggesting the presence of denser material surrounding the central region. Further analysis using the partial covering model (\texttt{pcfabs}) reveals that the covering fraction remains close to unity in most epochs, indicating that the X-ray emitting region is largely obscured. Slight deviations in the covering factor hint at inhomogeneities in the absorbing medium. The use of the ionized absorber model (\texttt{zxipcf}) shows consistently low ionization parameters $(\log \xi \sim -1.3 ~\text{to}~-1.7)$, indicating a persistent warm absorber with moderate ionization in different epochs. We also tested whether the warm absorber could conceal a soft excess component. For the XMM2+Nu1 and XMM3+Nu2 datasets, including additional soft component in the model did not significantly improve the spectral fit, and the parameters remained unconstrained. This suggests that any underlying soft excess is weak and largely masked by absorption features in the soft X-ray band.

These results collectively support a scenario in which Mrk~1040 is embedded in a complex, multi-phase medium that evolves in response to variations in the accretion rate and coronal structure. The absorber may locate near the torus, the broad-line region (BLR), or even the inner disc. Variability in $N_{\rm H}$ can result from clumpy clouds crossing the line of sight \citep{Risaliti2007, Markowitz2014}, or from dynamic processes such as thermally \citep{Begelman1983, Krongold2007} or magnetically driven outflows \citep{Fukumura2015, Braito2018} . The observed anti-correlation between $N_{\rm H}$ and soft excess luminosity ($L_{\rm SE}$) suggests that enhanced radiation pressure during high-luminosity states may either ionize or displace absorbing material \citep{Mehdipour2017, Ricci2010}. Alternatively, changes in $N_{\rm H}$ may also reflect evolving ionization states of a partially ionized warm medium \citep{Ebrero2016, Mizumoto2017}. Although we are unable to constrain the inclination angle, the presence of variable absorption and weak reflection features may indicate an intermediate viewing angle where both disc and outflow signatures can be observed. However, large uncertainties in $N_{\rm H}$, particularly in later epochs, limit our ability to draw firm conclusions about the evolution of the absorbing medium. Future observations with better spectral resolution will be critical to probing the geometry and physical nature of the absorbing medium. A detailed discussion of the impact of variable absorption on the lag behavior is given in next section (Section~\ref{sec:relation}).

\begin{table*}
\begin{center}
\caption{Variation of ${\rm N}_{\rm H}$ using different absorption models.}
\label{tab:nh}
\begin{tabular}{llccccccc}
\hline
Model & Parameter & XMM1 & Su & XMM2+Nu1 & XMM3+Nu2 & XRTc & XRT21 & XRT24 \\
\hline
\texttt{zTbabs} & ${\rm N}_{\rm H}~(10^{21}\text{cm}^{-2})$ & $1.71^{+0.74}_{-0.75}$ & $0.65^{+0.07}_{-0.03}$ & $1.25^{+0.19}_{-0.26}$ & $1.09^{+0.20}_{-0.22}$ & $2.28^{+0.29}_{-0.30}$ & $2.37^{+3.55}_{-1.74}$ & $2.26^{+0.89}_{-0.87}$ \\
\hline
\texttt{pcfabs} & ${\rm N}_{\rm H}~(10^{21}\text{cm}^{-2})$ & $1.69^{+0.66}_{-0.67}$ & $0.61^{+0.07}_{-0.08}$ & $1.33^{+0.22}_{-0.23}$ & $1.27^{+0.20}_{-0.21}$ & $2.89^{+0.40}_{-0.39}$ & $3.71^{+0.94}_{-0.97}$ & $4.29^{+0.89}_{-0.90}$ \\
&  &  &  &  &  &  &  &  \\
  & $ C_f^\dagger$ & $<1.00$ & $0.96^{+0.02}_{-0.23}$ & $<1.00$ & $1.00^f$ & $<1.00$ & $0.81^{+0.19}_{-0.18}$ & $0.91^{+0.86}_{-0.85}$ \\

\hline
\texttt{zxipcf} & ${\rm N}_{\rm H}~(10^{21}\text{cm}^{-2})$ & $1.53^{+0.06}_{-0.04}$ & $0.54^{+0.02}_{-0.04}$ & $1.08^{+0.24}_{-0.24}$ & $1.09^{+0.13}_{-0.14}$ & $2.25^{+0.25}_{-0.39}$ & $2.93^{+0.18}_{-0.10}$ & $3.86^{+0.12}_{-0.15}$ \\
&  &  &  &  &  &  &  &  \\
  & $C_f^\dagger$ & $<1.00$ & $0.94^{+0.02}_{-0.23}$ & $<1.00$ & $<1.00$ & $<1.00$ & $0.95^{+0.19}_{-0.18}$ & $0.87^{+0.86}_{-0.85}$ \\
&  &  &  &  &  &  &  &  \\
  & $\log \xi$ & $-1.36^{+0.03}_{-0.05}$ & $-1.33^{+0.02}_{-0.03}$ & $-1.37^{+0.04}_{-0.02}$ & $-1.36^{+0.03}_{-0.03}$ & $-1.35^{+0.04}_{-0.02}$ & $-1.69^{+1.10}_{-1.18}$ & $-1.60^{+1.57}_{-1.39}$ \\

\hline
\hline
\end{tabular}
\end{center}
\flushleft \textbf{Note:} $^\dagger$ Covering factor. $f$ indicates a frozen parameter.
\end{table*}

\subsection{Relation between different parameters}
\label{sec:relation}
\begin{figure}
\begin{center}
\includegraphics[trim={0 1.5cm 0cm 0},scale=0.8]{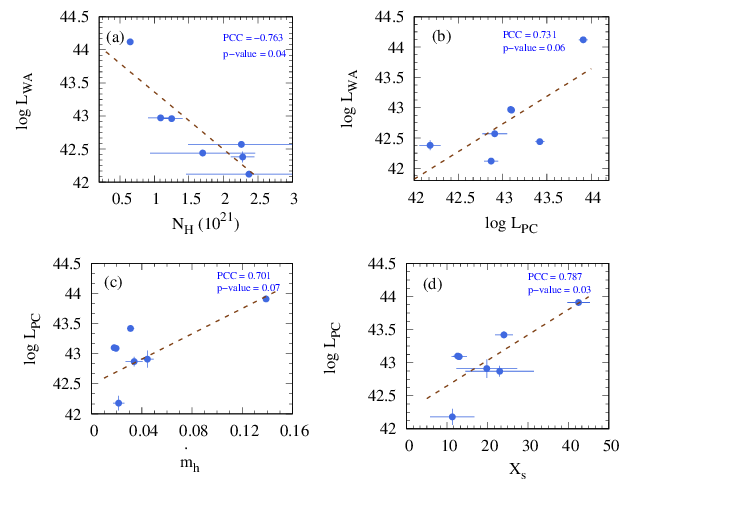}
\caption{Correlation between different parameters. The corresponding values of each parameter are given in Section~\ref{sec:spec}.}
\label{fig:corr}
\end{center}
\end{figure}

In this subsection, we investigate the interdependence of various observed and derived spectral parameters for Mrk~1040, based on multi-epoch observations over the 15-year period from 2009 to 2024 (see Section~\ref{sec:spec}). Although each spectral and temporal parameter individually contributes to the observed variability, they are not completely independent. The correlations between different parameters can provide deeper insight into the underlying physical processes driving the observed variability. We used Pearson Correlation Coefficient\footnote{\url{https://www.socscistatistics.com/tests/pearson/default2.aspx}} (PCC) to calculate the degree of correlation between the spectral parameters. The correlations, along with the statistical significance, are presented in Figure~\ref{fig:corr}. Due to the limited number of observations (seven), it is challenging to draw strong statistical conclusions about the correlations. However, we did notice a trend in the correlation between the parameters over this 15-year time frame, which can be understood through physical interpretation. From the phenomenological spectral modeling of Mrk~1040, we find that both the intrinsic hydrogen column density ($N_{\rm H}$) and the X-ray luminosities ($L_{x}$) exhibit temporal variability (see Section~\ref{sec:nh}). Since low-energy absorption primarily affects the soft X-ray band (below 3 keV), we further investigate the correlation between $N_{\rm H}$ and the soft excess luminosity ($L_{WA}$). A significant anti-correlation is observed between $N_{\rm H}$ and $L_{WA}$, with a PCC of -0.763 and the corresponding $p$-value of 0.04 (Figure~\ref{fig:corr} (a) ). This indicates that as the soft excess luminosity increases, the enhanced radiation pressure may either push the absorbing material outward or ionize it more effectively, resulting in a decrease in line-of-sight column density. Such anti-correlation has also been reported in previous studies of obscured AGNs, where changes in the luminosity affect the distribution of the intrinsic absorber \citep{Akylas2008, Fabian2009}. Furthermore, we investigated the correlation between the soft X-ray luminosity ($L_{WA}$) and the primary continuum luminosity ($L_{PC}$). We found a strong positive correlation (PCC = 0.731, $p$-value = 0.06) between $L_{WA}$ and $L_{PC}$, which is consistent with the trend commonly observed in Seyfert~1 AGNs \citep{Nandi2021, Nandi2023, Nandi2024}. This positive correlation suggests a common physical origin for both the soft and hard X-ray bands \citep{Waddell2020, Nandi2021, Nandi2024}.

To further explore the connection between the accretion flow dynamics and hard X-ray emission, we investigate the correlation between the sub-Keplerian (halo) accretion rate ($\dot{m}_h$) and the primary continuum luminosity ($L_{PC}$). We found a positive correlation between $\dot{m}_h$ and $L_{PC}$ with PCC of 0.701 and $p$-value of 0.07 (Figure~\ref{fig:corr} (c)). In the TCAF framework, the Compton cloud responsible for hard X-ray emission is predominantly fed by the halo component. As $\dot{m}_h$ increases, more low-angular-momentum matter is supplied to the post-shock region, commonly referred to as the Compton cloud. This leads to a larger and denser Comptonizing region, which enhances the inverse Compton scattering of soft photons and results in increased hard X-ray luminosity \citep{Nandi2012}. The observed positive correlation between $\dot{m}_h$, the properties of the Compton cloud, and the X-ray luminosity provides direct observational support for the two-component advective flow (TCAF) model, linking the accretion dynamics with the emergent X-ray emission.

We find a positive trend (PCC = 0.787 and $p$-value = 0.03) between the estimated coronal size (shock location $X_s$) and the X-ray luminosity $L_X$ (Figure~\ref{fig:corr}d). In two-component advective flow framework an increase in accretion rate can lead to an expanded Comptonizing region and enhanced hard X-ray emission \citep{CT95}, qualitatively consistent with such a trend. However, given the small number of epochs, this relation is tentative and must be interpreted with caution. In particular, variations in $X_s$ and $L_X$ may both be driven by a common underlying parameter, such as the accretion rate, rather than implying a direct causal connection. We further note that Comptonization models predict soft–hard delays of order tens of ks for corona extending to several tens of $r_g$ \citep{ST80}, but these delays can be partially or fully compensated by light-travel effects, producing near-zero net lags in observations \citep{Uttley2014, Mastroserio2020, Nandi2021}. We thus regard the $X_s$–$L_X$ relation as an intriguing but inconclusive indication of coronal evolution, which will require more temporal sampling and quantitative modeling, such as propagating-fluctuation or reverberation frameworks\citep{Kara2016}, to be robustly established.
\section{Conclusions}
We present a long-term ($\sim$15 years) multi-epoch spectral and timing study of Mrk~1040. We confirm the presence of the strong soft excess in the 2009 XMM1 observation, consistent with the earlier report by \citet{Tripathi2011}. Our work extends this result by performing a uniform, physically motivated modeling of all available observations (2009--2024) to quantify the temporal evolution of the soft excess, coronal geometry, and absorber properties. These variations indicate that Mrk~1040 likely undergoes transitions between different accretion states. Our findings offer new insights into the long-term physical evolution of the accretion flow and circumnuclear environment around the central supermassive black hole in this AGN. The key results are summarized below:

\begin{enumerate}
    \item[1.] Timing analysis of the 2009 observation confirms the presence of a strong soft-excess component, consistent with the earlier findings of \citep{Tripathi2011}, and this is further supported by our spectral modeling. The excess emission is well described by a warm and extended Comptonizing corona with an electron temperature of $kT_{\rm e,warm} \sim 0.26$ keV and a radial extent of $R_{\rm warm} \sim 30~r_g$. In subsequent epochs, this soft excess is no longer statistically significant, suggesting a substantial change in the accretion geometry or radiative properties. Its disappearance may result from multiple factors, such as enhanced ionized absorption, intrinsic weakening of the warm component, or possible inner-disc truncation, rather than truncation alone. The timing analysis further indicates that the corona became more stable during the later observations.

    \item[2.] Both spectral and timing analyses reveal significant variability in the Keplerian and sub-Keplerian accretion rates over time. The evolution from a multi-zone corona in 2009 to a more compact and thermally stable corona from 2015 onward suggests a shift between radiatively efficient and inefficient accretion modes, consistent with dynamic disc–corona coupling.

    \item[3.]  The Fe K$\alpha$ emission line evolves notably across epochs. It gets weaker during periods of coronal expansion and disc truncation (2013) and becomes stronger when the corona is compact (2015). The non-detection of this emission line feature from 2021 to 2024 is attributed to instrumental limitations rather than physical disappearance. The presence of the line in earlier epochs points to partial illumination of distant material such as the outer disc.

    \item[4.]  A trend of correlations is found between several key physical parameters. The increase in X-ray luminosity is strongly associated with increases in both the accretion rate $(\dot{m}_h)$ and the Compton cloud size, indicating a strong disc–corona interaction governed by evolving inflow dynamics. The enhanced luminosity is likely associated with the radiation pressure that displaces the absorbing medium located along the line of sight. This absorber exhibits notable temporal variation: the lowest column density ($N_{\rm H}$) was recorded in 2013 during a high-luminosity and low-obscuration phase, while later epochs reveal elevated $N_{\rm H}$ values, indicative of a denser, possibly clumpy medium. However, the large uncertainties in the $N_{\rm H}$ measurements, especially in the \textit{Swift}/XRT observations, limit firm conclusions about its long-term evolution. Future high-quality, high-resolution observations will be essential to better constrain the nature of the absorber.

    \item[5.] Using physical spectral models such as \texttt{TCAF}, we estimate the mass of the central black hole to be $M_{BH} = (4.50 \pm 1.62) \times 10^7~M_\odot$.
  
\end{enumerate}
\section*{Acknowledgments}
We sincerely thank the reviewer for providing insightful comments and constructive suggestions that helped us improve the manuscript. The research work of PN and SKC is conducted at the Indian Centre for Space Physics (ICSP), Kolkata, West Bengal, India. The research work in the Physical Research Laboratory, Ahmedabad, is funded by the Department of Space, Government of India. The data and the software used in this work are taken from the High Energy Astrophysics Science Archive Research Center (HEASARC), which is a service of the Astrophysics Science Division at NASA/GSFC and the High Energy Astrophysics Division of the Smithsonian Astrophysical Observatory. This work has made use of data obtained from the {\it Nustar} mission, a project led by Caltech, funded by NASA and managed by NASA/JPL, and has utilized the NuSTARDAS software package, jointly developed by the ASDC, Italy and Caltech, USA. This work made use of data {\it Swift} supplied by the UK Swift Science Data Centre at the University of Leicester. We acknowledge the use of public data from the Swift data archive. This research has used observations obtained with {\it XMM-Newton}, an ESA science mission with instruments and contributions directly funded by ESA Member States and NASA.

\section*{Data Availability}
We used archival data of {\it Swift}/XRT, {\it XMM-Newton}, {\it Suzaku} and {\it NuSTAR} observatories for this work. These data are publicly available on their corresponding websites. Appropriate links are given in the text.



\bibliographystyle{mnras}
\bibliography{Mrk1040} 



\appendix

\section{Some additional material}
\begin{figure*}
\centering
\includegraphics[width=1.05\textwidth]{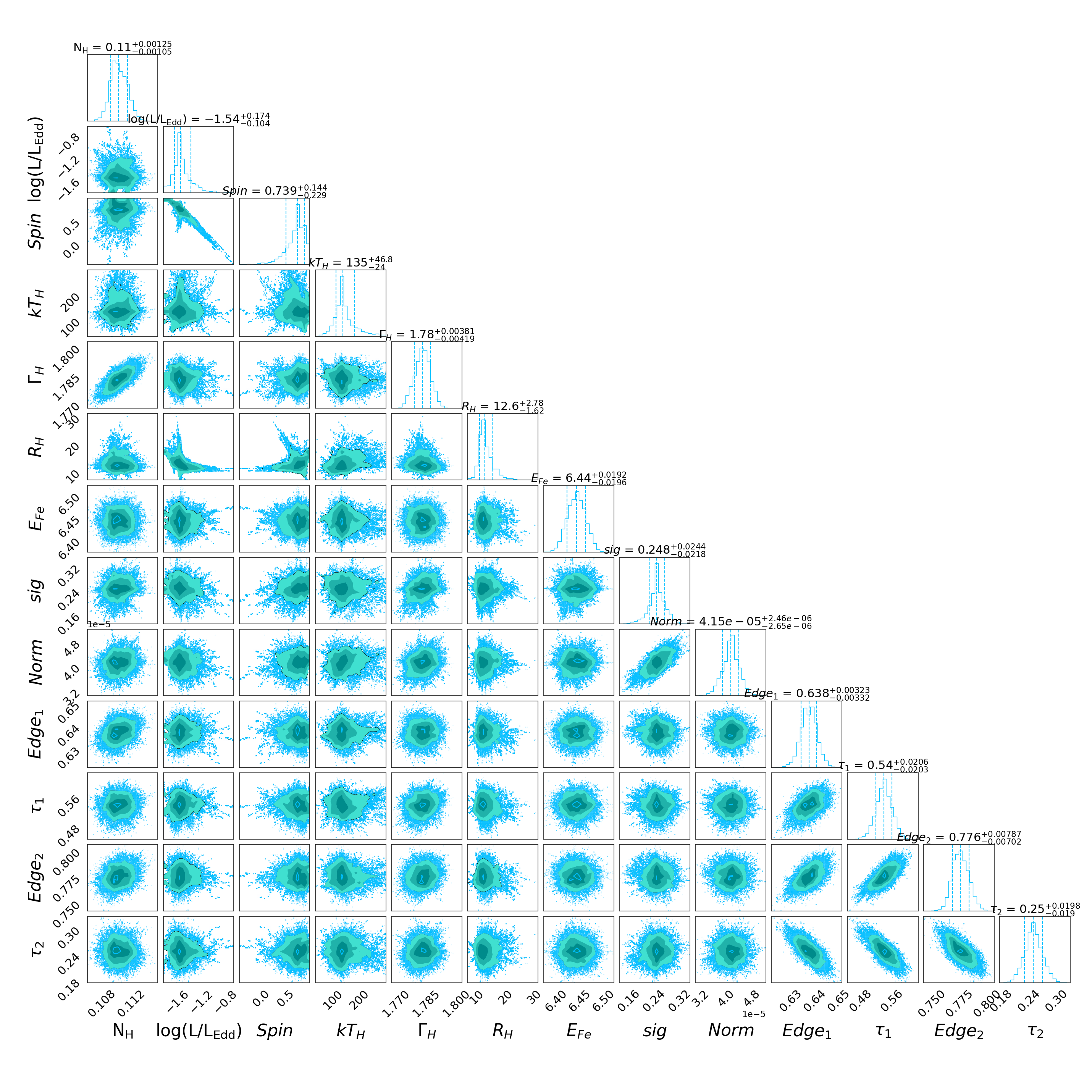}
\caption{Corner plots of spectral parameters from MCMC analysis for the XMM3+Nu2 observation using \texttt{AGNSED} model. 1D histograms represent the probability distribution. Three vertical lines in the 1D distribution show 16\%, 50\%, and 90\% quintiles. We used \texttt{CORNER.PY} \citep{Foreman2017} to plot the distributions.}
\label{fig:mcmc_a}
\end{figure*}

\begin{figure*}
\centering
\includegraphics[width=1.05\textwidth]{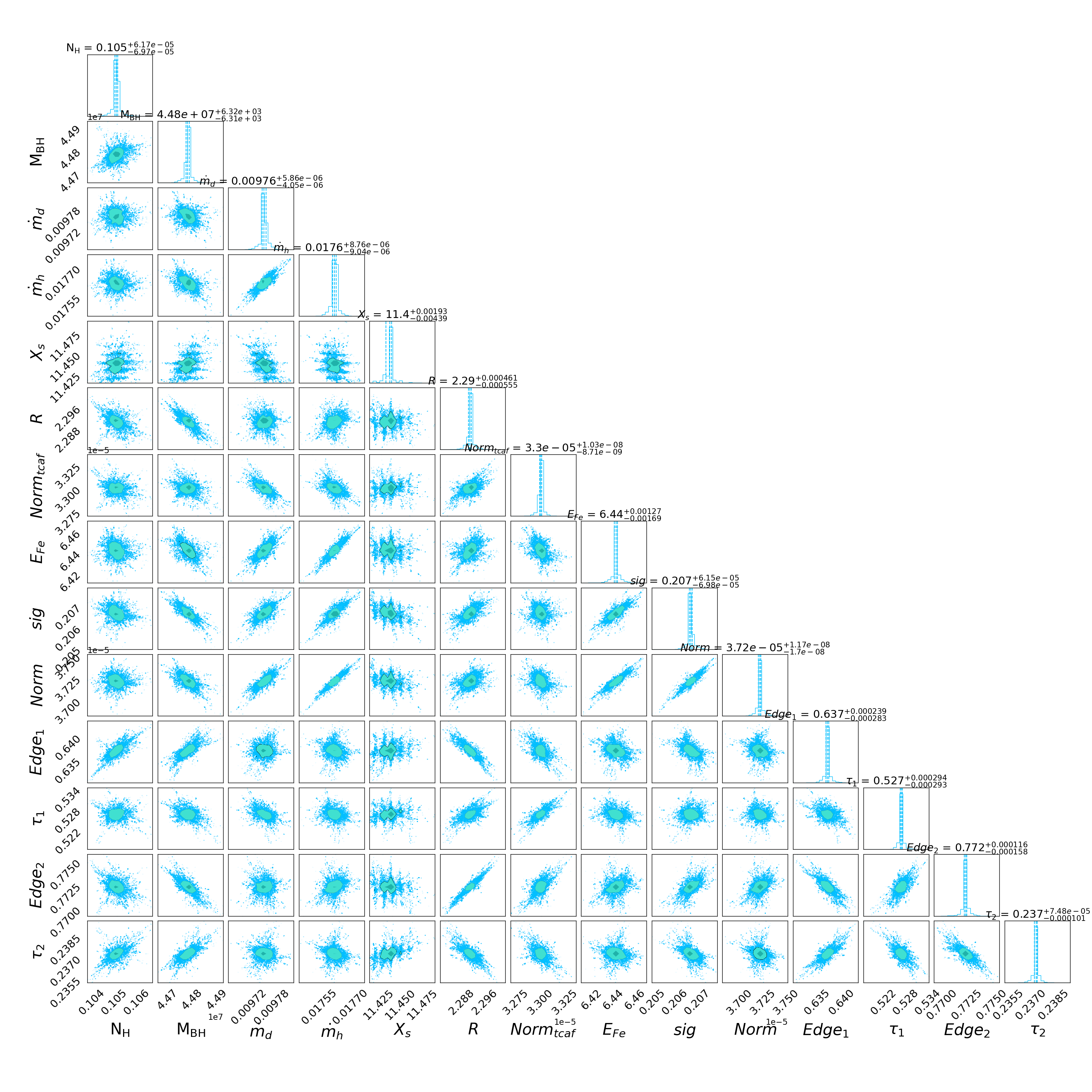}
\caption{Corner plots of spectral parameters from MCMC analysis for the XMM3+Nu2 observation using \texttt{TCAF} model. 1D histograms represent the probability distribution. Three vertical lines in the 1D distribution show 16\%, 50\%, and 90\% quintiles. We used \texttt{CORNER.PY} \citep{Foreman2017} to plot the distributions.}
\label{fig:mcmc}
\end{figure*}


\begin{table}
\centering
\caption{Variability in different energy bands was analyzed using 500s (0.5 ks) binned light curves from {\it XMM-Newton} and {\it NuSTAR}. We applied the \texttt{ZDCF} and \texttt{ICF} to calculate the correlations between bands $\Delta E_1$ and $\Delta E_2$. Correlation peaks were obtained by Gaussian fitting, with lag uncertainties ($\epsilon$) estimated via \citet{Gaskell1987}. If $\epsilon$ was smaller than the bin size, the bin size was adopted as the error.}
\label{tab:corr}
\begin{tabular}{lc@{\hspace{3pt}}c@{\hspace{6pt}}c@{\hspace{6pt}}c@{\hspace{6pt}}c}
\hline
ID & $\Delta E_1$ & $\Delta E_2$ & Peak & $\epsilon$ & $\tau$ \\
 & (keV) & (keV) &  & (ks) & (ks) \\
\hline
XMM1 & 3.0--4.0 & 0.3--0.5 & $0.52\pm0.25$ & $0.63$ & $0.47\pm0.63$  \\
     & 3.0--4.0 & 0.5--1.0 & $0.53\pm0.08$ & $0.96$ & $0.57\pm0.96$  \\
     & 3.0--4.0 & 1.0--2.0 & $0.68\pm0.06$ & $0.62$ & $0.67\pm0.62$  \\
     & 3.0--4.0 & 2.0--3.0 & $0.70\pm0.05$ & $0.67$ & $0.41\pm0.67$  \\
     & 3.0--4.0 & 4.0--5.0 & $0.58\pm0.15$ & $0.78$ & $0.75\pm0.78$  \\
     & 3.0--4.0 & 5.0--6.0 & $0.51\pm0.08$ & $0.78$ & $0.73\pm0.78$  \\
     & 3.0--4.0 & 6.0--7.0 & $0.51\pm0.16$ & $1.15$ & $0.95\pm1.15$  \\
     & 3.0--4.0 & 7.0--8.0 & $0.27\pm0.22$ & $1.39$ & $0.52\pm1.39$  \\  
     & 3.0--4.0 & 8.0--10.0& $0.42\pm0.29$ & $0.83$ & $0.49\pm0.83$  \\ 
     &  &  &  &  &  \\
XMM2 & 3.0--4.0 & 0.3--0.5 & $0.68\pm0.06$ & $0.81$ & $0.31\pm0.81$  \\
     & 3.0--4.0 & 0.5--1.0 & $0.75\pm0.04$ & $0.78$ & $0.73\pm0.78$  \\
     & 3.0--4.0 & 1.0--2.0 & $0.76\pm0.04$ & $0.73$ & $0.72\pm0.73$  \\
     & 3.0--4.0 & 2.0--3.0 & $0.70\pm0.04$ & $0.77$ & $0.53\pm0.77$  \\
     & 3.0--4.0 & 4.0--5.0 & $0.63\pm0.06$ & $1.07$ & $1.03\pm1.07$  \\
     & 3.0--4.0 & 5.0--6.0 & $0.54\pm0.07$ & $1.19$ & $1.04\pm1.19$  \\
     & 3.0--4.0 & 6.0--7.0 & $0.64\pm0.17$ & $2.13$ & $-1.79\pm2.13$  \\
     & 3.0--4.0 & 7.0--8.0 & $0.49\pm0.19$ & $2.96$ & $2.36\pm2.96$  \\  
     & 3.0--4.0 & 8.0--10.0& $0.44\pm0.19$ & $1.67$ & $5.70\pm1.67$  \\ 
     &  &  &  &  &  \\
XMM3 & 3.0--4.0 & 0.3--0.5 & $0.70\pm0.18$ & $1.23$ & $-0.47\pm1.23$  \\
     & 3.0--4.0 & 0.5--1.0 & $0.63\pm0.07$ & $0.96$ & $-0.47\pm0.96$  \\
     & 3.0--4.0 & 1.0--2.0 & $0.72\pm0.06$ & $0.85$ & $-0.52\pm0.95$  \\
     & 3.0--4.0 & 2.0--3.0 & $0.74\pm0.05$ & $0.69$ & $0.39\pm0.69$  \\
     & 3.0--4.0 & 4.0--5.0 & $0.63\pm0.05$ & $0.92$ & $0.36\pm0.92$  \\
     & 3.0--4.0 & 5.0--6.0 & $0.70\pm0.13$ & $0.84$ & $0.99\pm0.84$  \\
     & 3.0--4.0 & 6.0--7.0 & $0.50\pm0.28$ & $1.45$ & $0.32\pm1.45$  \\
     & 3.0--4.0 & 7.0--8.0 & $0.53\pm0.21$ & $1.39$ & $-0.93\pm1.39$  \\ 
     & 3.0--4.0 & 8.0--10.0& $0.49\pm0.18$ & $1.47$ & $1.16\pm1.47$  \\ 
     &  &  &  &  &  \\
Nu1  & 3.0--4.0 & 4.0--5.0 & $0.44\pm0.22$ & $1.22$ & $-5.01\pm1.22$  \\
     & 3.0--4.0 & 5.0--6.0 & -- & -- & --  \\
     & 3.0--4.0 & 6.0--7.0 & -- & -- & --  \\
     & 3.0--4.0 & 7.0--8.0 & -- & -- & --  \\ 
     & 3.0--4.0 & 8.0--10.0& -- & -- & --  \\
     & 3.0--4.0 &10.0--15.0& -- & -- & --  \\
     & 3.0--4.0 &15.0--20.0& -- & -- & --  \\
     & 3.0--4.0 &20.0--30.0& -- & -- & --  \\
     & 3.0--4.0 &30.0--40.0& -- & -- & --  \\  
     &  &  &  &  &  \\
Nu2  & 3.0--4.0 & 4.0--5.0 & -- & -- & --  \\
     & 3.0--4.0 & 5.0--6.0 & -- & -- & --  \\
     & 3.0--4.0 & 6.0--7.0 & -- & -- & --  \\
     & 3.0--4.0 & 7.0--8.0 & -- & -- & --  \\ 
     & 3.0--4.0 & 8.0--10.0& -- & -- & --  \\
     & 3.0--4.0 &10.0--15.0& -- & -- & --  \\
     & 3.0--4.0 &15.0--20.0& -- & -- & --  \\
     & 3.0--4.0 &20.0--30.0& -- & -- & --  \\
     & 3.0--4.0 &30.0--40.0& -- & -- & --  \\  
     &  &  &  &  &  \\

\hline
\end{tabular}
\end{table}

\begin{table}
\centering
\caption{Variability statistics in different energy bands using 500~s binned light curves from {\it XMM-Newton} and {\it NuSTAR} observations.}
\label{tab:fvar}
\begin{tabular}{lc@{\hspace{3pt}}c@{\hspace{6pt}}c@{\hspace{6pt}}c@{\hspace{6pt}}c@{\hspace{6pt}}c@{\hspace{6pt}}c}
\hline
ID & Energy & $N$ & $x_{\max}$ & $x_{\min}$ & $\mu$ & $\sigma^2_{\rm NXS}$ & $F_{\rm var}$ \\
 & (keV) &  & (cts/s) & (cts/s) & (cts/s) & ($10^{-2}$) & (\%) \\
\hline
XMM1 & 0.3--0.5 & 177 & 1.12 & 0.52 & 0.75 & $2.93 \pm 0.16$ & $17.1 \pm 1.0$ \\
     & 0.5--1.0 & 177 & 2.11 & 1.11 & 1.50 & $2.01 \pm 0.09$ & $14.2 \pm 0.8$ \\
     & 1.0--2.0 & 177 & 3.75 & 2.14 & 2.74 & $1.31 \pm 0.05$ & $11.4 \pm 0.6$ \\
     & 2.0--3.0 & 177 & 1.27 & 0.65 & 0.91 & $0.99 \pm 0.09$ & $9.9 \pm 0.7$ \\
     & 3.0--4.0 & 177 & 0.67 & 0.32 & 0.46 & $0.88 \pm 0.13$ & $9.4 \pm 0.8$ \\
      & 4.0--5.0 & 177 & 0.41 & 0.21 & 0.30 & $0.52 \pm 0.14$ & $7.1 \pm 1.0$ \\
      & 5.0--6.0 & 177 & 0.27 & 0.14 & 0.20 & $0.35 \pm 0.18$ & $5.9 \pm 1.6$ \\
      & 6.0--7.0 & 177 & 0.21 & 0.07 & 0.16 & $0.25 \pm 0.22$ & $5.0 \pm 2.2$ \\
      & 7.0--8.0 & 177 & 0.18 & 0.04 & 0.08 & $0.71 \pm 0.47$ & $8.4 \pm 2.9$ \\
      & 8.0--10.0& 177 & 0.14 & 0.03 & 0.07 & $1.23 \pm 0.57$ & $11.0 \pm 2.6$ \\
      &  &  &  &  &  &  &  \\
XMM2 & 0.3--0.5 & 181 & 0.73 & 0.30 & 0.56 & $1.31 \pm 0.13$ & $11.47 \pm 0.81$ \\
     & 0.5--1.0 & 181 & 1.47 & 0.79 & 1.14 & $1.28 \pm 0.08$ & $11.5 \pm 0.8$ \\
     & 1.0--2.0 & 181 & 2.80 & 1.64 & 2.23 & $1.14 \pm 0.05$ & $10.70 \pm 0.62$ \\
     & 2.0--3.0 & 181 & 1.04 & 0.52 & 0.78 & $0.97 \pm 0.09$ & $9.85 \pm 0.70$ \\
     & 3.0--4.0 & 181 & 0.54 & 0.29 & 0.41 & $0.76 \pm 0.12$ & $8.72 \pm 0.84$ \\    
     & 4.0--5.0 & 181 & 0.38 & 0.19 & 0.28 & $0.72 \pm 0.16$ & $8.51 \pm 1.04$ \\    
     & 5.0--6.0 & 181 & 0.26 & 0.12 & 0.19 & $0.62 \pm 0.20$ & $7.85 \pm 1.35$ \\    
     & 6.0--7.0 & 181 & 0.22 & 0.10 & 0.15 & $0.27 \pm 0.21$ & $5.18 \pm 2.08$ \\    
     & 7.0--8.0 & 181 & 0.12 & 0.04 & 0.08 & $0.25 \pm 0.38$ & $4.98 \pm 3.84$ \\    
     & 8.0--10.0& 181 & 0.11 & 0.04 & 0.07 & $7.39 \pm 0.44$ & $7.75 \pm 1.89$ \\    
     &  &  &  &  &  &  &  \\
Nu1 & 3.0--4.0 & 100 & 0.05 & 0.01 & 0.03 & $1.47 \pm 1.52$ & $12.14 \pm 6.34$ \\
     & 4.0--5.0 & 100 & 0.38 & 0.24 & 0.29 & $0.23 \pm 0.17$ & $4.81 \pm 6.34$ \\ 
     & 5.0--6.0 & 100 & 0.21 & 0.11 & 0.16 & $0.40 \pm 0.32$ & $6.34 \pm 2.56$ \\ 
     & 6.0--7.0 & 100 & 0.20 & 0.10 & 0.15 & $0.31 \pm 0.32$ & $5.54 \pm 2.93$ \\ 
     & 7.0--8.0 & 100 & 0.16 & 0.06 & 0.11 & $0.64 \pm 0.46$ & $8.03 \pm 2.90$ \\ 
     & 8.0--10.0& 100 & 0.22 & 0.12 & 0.16 & $0.10 \pm 0.27$ & $3.19 \pm 4.20$ \\ 
     &10.0--15.0& 100 & 0.26 & 0.14 & 0.19 & $0.76 \pm 0.30$ & $8.70 \pm 1.85$ \\ 
     &15.0--20.0& 100 & 0.14 & 0.05 & 0.08 & $0.20 \pm 0.53$ & $4.48 \pm 5.91$ \\ 
     &20.0--30.0& 100 & 0.11 & 0.03 & 0.06 & $1.16 \pm 0.79$ & $1.08 \pm 3.73$ \\ 
     &30.0--40.0& 100 & 0.05 & 0.01 & 0.02 & $1.85 \pm 1.98$ & $13.61 \pm 0.73$ \\ 
     &  &  &  &  &  &  &  \\
XMM3 & 0.3--0.5 & 162 & 0.92 & 0.46 & 0.69 & $1.50 \pm 0.12$ & $12.26 \pm 0.85$ \\
     & 0.5--1.0 & 162 & 1.70 & 0.93 & 1.32 & $1.21 \pm 0.08$ & $11.17 \pm 0.72$ \\
     & 2.0--3.0 & 162 & 0.99 & 0.56 & 0.78 & $1.10 \pm 0.10$ & $10.49 \pm 0.76$ \\
     & 3.0--4.0 & 162 & 0.53 & 0.26 & 0.39 & $1.01 \pm 0.14$ & $9.95 \pm 0.92$ \\
     & 4.0--5.0 & 162 & 0.35 & 0.18 & 0.26 & $0.93 \pm 0.19$ & $9.45 \pm 1.13$ \\
     & 5.0--6.0 & 162 & 0.28 & 0.11 & 0.18 & $1.41 \pm 0.28$ & $11.63 \pm 1.36$ \\
     & 6.0--7.0 & 162 & 0.20 & 0.08 & 0.14 & $0.54 \pm 0.26$ & $6.62 \pm 2.01$ \\
     & 7.0--8.0 & 162 & 0.11 & 0.04 & 0.07 & $0.39 \pm 0.45$ & $5.99 \pm 3.78$ \\
     & 8.0--10.0& 162 & 0.10 & 0.03 & 0.07 & $1.41 \pm 0.60$ & $11.73 \pm 2.62$ \\
     &  &  &  &  &  &  &  \\
Nu2 & 3.0--4.0 & 108 & 0.06 & 0.01 & 0.03 & $0.35 \pm 1.34$ & $5.93 \pm 11.27$ \\
     & 4.0--5.0 & 108 & 0.36 & 0.19 & 0.26 & $0.66 \pm 0.22$ & $8.13 \pm 1.49$ \\
     & 5.0--6.0 & 108 & 0.21 & 0.08 & 0.15 & $1.39 \pm 0.42$ & $11.81 \pm 1.95$ \\
     & 6.0--7.0 & 108 & 0.18 & 0.07 & 0.14 & $0.21 \pm 0.32$ & $4.55 \pm 3.54$ \\
     & 7.0--8.0 & 108 & 0.14 & 0.05 & 0.10 & $0.21 \pm 0.45$ & $4.59 \pm 4.97$ \\
     & 8.0--10.0& 108 & 0.19 & 0.08 & 0.15 & $0.41 \pm 0.34$ & $6.38 \pm 2.68$ \\
     &10.0--15.0& 108 & 0.25 & 0.12 & 0.17 & $0.65 \pm 0.33$ & $8.09 \pm 2.08$ \\
     &15.0--20.0& 108 & 0.12 & 0.04 & 0.07 & $1.47 \pm 0.71$ & $12.14 \pm 7.10$ \\
     &20.0--30.0& 108 & 0.12 & 0.03 & 0.06 & $1.90 \pm 0.89$ & $13.8 \pm 7.58$ \\
     &30.0--40.0& 108 & 0.05 & 0.01 & 0.02 & $1.91 \pm 2.08$ & $13.9 \pm 7.59$ \\
\hline
\end{tabular}
\end{table}


\bsp	
\label{lastpage}
\end{document}